\documentclass[apj]{emulateapj}
\usepackage{amssymb,latexsym,amsmath}
\usepackage{apjfonts}
\usepackage{graphicx}
\usepackage{natbib}
\usepackage{bbm}
\def \arcmin {\hbox{$^\prime$}}
\def \arcsec {\hbox{$^{\prime\prime}$}}

\def\spose#1{\hbox to 0pt{#1\hss}}
\def\ltsim{$\mathrel{\spose{\lower 3pt\hbox{$\sim$}}
        \raise 2.0pt\hbox{$<$}}$\thinspace}
\def\gtsim{$\mathrel{\spose{\lower 3pt\hbox{$\sim$}}
        \raise 2.0pt\hbox{$>$}}$\thinspace}

\newcommand{\apec}{APEC}

\newcommand{\zsun }{${\rm Z_\odot}$}

\newcommand{\chandra }{{\em Chandra}}

\newcommand{\xspec }{{\em Xspec}}

\newcommand{\fxunits}{\mbox{erg cm$^{-2}$ s$^{-1}$}}

\newcommand{\rxte }{{\em RXTE}}
\newcommand{\xmm }{{\em XMM}}
\newcommand{\asca }{{\em ASCA}}
\newcommand{\rosat }{{\em ROSAT}}

\newcommand{\sax }{{\em Beppo-SAX}}
\newcommand{\nustar }{{\em NuSTAR}}
\newcommand{\suzaku }{{\em Suzaku}}
\newcommand{\swift }{{\em Swift}}

\newcommand\omegam{\hbox{{$\Omega_{\rm m}$}}}
\newcommand\omegalambda{\hbox{{$\Omega_{\Lambda}$}}}
\newcommand\kmsmpc{{\rm km s$^{-1}$ Mpc$^{-1}$}}
\newcommand\ho{\hbox{{$H_{0}$}}}

\slugcomment{Received 2014 November 4; accepted 2014 December 28; published 2015 ???}
\shorttitle{A \nustar\ observation of the center of Coma}
\shortauthors{Gastaldello et~al.}
\begin{document}
\title{A \nustar\ observation of the center of the Coma Cluster}
\author {Fabio Gastaldello,\altaffilmark{1,2}
Daniel R. Wik,\altaffilmark{3,4}
S.~Molendi,\altaffilmark{1}
N.~J.~Westergaard,\altaffilmark{5}
A.~Hornstrup,\altaffilmark{5}
G.~Madejski,\altaffilmark{6}
D.~D.~M.~Ferreira,\altaffilmark{5}
S.~E.~Boggs,\altaffilmark{7}
F.~E.~Christensen,\altaffilmark{5}
W.~W.~Craig,\altaffilmark{7,8}
B.~W.~Grefenstette,\altaffilmark{9}
C.~J.~Hailey,\altaffilmark{10}
F.~A.~Harrison,\altaffilmark{9}
K.~K.~Madsen,\altaffilmark{9}
D.~Stern,\altaffilmark{11}
W.~W.~Zhang\altaffilmark{3}
}
\altaffiltext{1}{IASF-Milano, INAF, Via Bassini 15, I-20133 Milan, Italy; gasta@lambrate.inaf.it}
\altaffiltext{2}{Department of Physics and Astronomy, University of
California at Irvine, 4129 Frederick Reines Hall, Irvine, CA 92697-4575, USA}
\altaffiltext{3}{Astrophysics Science Division, 
NASA/Goddard Space Flight Center,
Greenbelt, MD 20771, USA}
\altaffiltext{4}{Department of Physics and Astronomy, 
Johns Hopkins University, Baltimore, MD 21218, USA}
\altaffiltext{5}{ DTU Space, National Space Institute, Technical University of Denmark, 
Elektrovej 327, DK-2800 Lyngby, Denmark }
\altaffiltext{6}{ Kavli Institute for Particle Astrophysics and Cosmology, 
SLAC National Accelerator Laboratory, Menlo Park, CA 94025, USA }
\altaffiltext{7}{Space Sciences Laboratory, University of California, Berkeley, CA 94720, USA }
\altaffiltext{8}{Lawrence Livermore National Laboratory, Livermore, CA 94550, USA }
\altaffiltext{9}{Cahill Center for Astronomy and Astrophysics, 
California Institute of Technology, Pasadena, CA 91125, USA }
\altaffiltext{10}{ Columbia Astrophysics Laboratory, Columbia University, 
New York, NY 10027, USA }
\altaffiltext{11}{Jet Propulsion Laboratory, California Institute of Technology, 
Pasadena, CA 91109, USA }

\begin{abstract}
We present the results of a 55ks \nustar\ observation of the core of the Coma Cluster. The global spectrum can be explained by 
thermal gas emission, with a conservative 90\% upper limit to non-thermal inverse Compton (IC) emission of $5.1 \times 10^{-12}$ \fxunits\ in a
12\arcmin\ $\times$ 12\arcmin\ field of view.
The brightness of the thermal component in this central region does not allow more stringent upper limits on the IC component when compared
with non-imaging instruments with much larger fields of view where claims of detections have been made.
Future mosaic \nustar\ observations of Coma will further address this issue.
The temperature map shows a relatively uniform temperature distribution with a gradient from the hot northwest side 
to the cooler southeast, in agreement with previous measurements. The temperature determination is robust
given the flat effective area and low background in the 3--20 keV band, making \nustar\ an ideal instrument to measure
high temperatures in the intracluster medium. 

\end{abstract}
\keywords{galaxies: clusters: general --- galaxies: clusters: individual (Coma) ---  X-rays: galaxies: clusters}
%
%
\section{Introduction} 
\label{Introduction} 
The Coma cluster is one of the best studied clusters of galaxies 
in the sky \citep[see][for an historical review]{Biviano:98}.
It has been explored at all wavelengths from radio to hard X-rays, and its
proximity, richness and brightness have been key for revealing new and unexpected phenomena,
such as radio halos and relics \citep[][and references therein]{Feretti.ea:12}.
It is one of the most spectacular examples of hierarchical structure formation, with 
strong evidence of its buildup by in-falling substructures found
in the galaxy distribution \citep[e.g.,][]{Colless.ea:96,Adami.ea:05,Adami.ea:09},
in the X-ray morphology and surface brightness variations 
\citep[e.g.,][]{Briel.ea:92,Vikhlinin.ea:94,Vikhlinin.ea:97,Neumann.ea:03,Andrade-Santos.ea:13,Sanders.ea:13b},
in the map of the Sunyaev-Zeldovich effect \citep{PlanckIntermediateX:13} and in its weak lensing reconstructed mass distribution \citep{Gavazzi.ea:09,Okabe.ea:10,Okabe.ea:14}.

The Coma cluster has been observed with virtually every X-ray observatory flown.
Detailed temperature maps of the large-scale emission of the cluster with mosaic observations  
using \asca\ \citep{Watanabe.ea:99}, 
\xmm\ \citep[e.g.,][]{Arnaud.ea:01,Neumann.ea:03,Schuecker.ea:04} and \suzaku\ \citep{Simionescu.ea:13}
reveal complex temperature variations indicative of recent mergers in this complex cluster. Examples are
the in-falling NGC 4839 group, a hot Western region, and cooler gas possibly associated 
with gas stripped from the in-falling group associated with NGC 4921 and NGC 4911. 
The temperature distribution in the central 10\arcmin\ around the two central galaxies, 
NGC 4889 and NGC 4874, is relatively homogeneous in the 8-10 keV range 
\citep[e.g.,][]{Arnaud.ea:01,Sato.ea:11} with a gradient from the hot northwest side
of the core to the cool ($\sim 7$ keV) southeast, the latter associated with linear, 
higher-density arms consisting of low-entropy material that was probably stripped 
from merging subclusters \citep{Sanders.ea:13b}.

In addition to the hot intra-cluster gas, which constitutes its main baryonic component, Coma hosts
a large scale magnetic field and relativistic electrons as revealed by the diffuse Mpc-scale synchrotron 
emission of the radio halo, the first and brightest radio halo (discovered by \citealt{Willson.ea:70}) 
and one of the best studied \citep[e.g.,][]{Giovannini.ea:93,Deiss.ea:97,Thierbach.ea:03,Brown.ea:11}.
For a collection of relativistic electrons, the total synchrotron luminosity depends both on the 
number of electrons and on the magnetic field $B$. However the same electrons will up-scatter 
cosmic microwave background (CMB) photons through inverse Compton (IC) interaction with a luminosity which will
depend on the number of electrons and the known energy density of the CMB. Therefore the measurement 
of an IC flux from a synchrotron source directly leads to a simultaneous determination of the average value of $B$
and the relativistic electron density \citep[e.g.,][]{Harris.ea:74}. The IC emission can in principle be observed at hard X-ray 
energies \citep{Rephaeli:77} because the exponential decline of the thermal bremsstrahlung continuum
is distinctly steeper than the expected non-thermal spectrum, potentially detectable as excess hard X-ray emission.
Coma was indeed the first object for which a detection of non-thermal emission was claimed 
\citep{Rephaeli.ea:99,Rephaeli.ea:02} based on data from  \rxte. \citet{Fusco-Femiano.ea:99,Fusco-Femiano.ea:04} also claimed a detection 
based on data from \sax. A number of claims of a hard X-ray excess have also been made in  
several other radio-halo clusters, although they are mostly of marginal significance \citep[see][for a review]{Rephaeli.ea:08}.
The \sax\ detection in Coma, which the most recent analysis puts at a confidence level of
4.8$\sigma$ with a non-thermal flux of $(1.30\pm0.40) \times 10^{-11}$ \fxunits\ in the 20-80 keV energy band \citep[e.g.,][]{Fusco-Femiano.ea:11}
has been very controversial \citep{Rossetti.ea:04,Fusco-Femiano.ea:07}. 

This contributed to further attempts to confirm IC emission at the claimed \rxte\ and \sax\ levels
with \suzaku\ and \swift, though these attempts have largely failed
\citep[see][for a review]{Ota:12}.  \citet{Wik.ea:09} performed a joint \xmm\ EPIC-pn and \suzaku\ HXD-PIN analysis of the Coma 
cluster and were unable to detect IC emission, finding an upper limit of $6.0 \times 10^{-12}$ \fxunits, 
2.5 times below the claimed \rxte\ and \sax\ detection. This discrepancy could still be resolved taking into account the smaller field of view (FOV) 
of the \suzaku\ HXD-PIN if the IC emission is very extended, beyond the radio halo \citep{Fusco-Femiano.ea:11}. 
\citet{Wik.ea:11} performed a joint \xmm\ EPIC-pn and \swift\ Burst Alert Telescope (BAT) analysis, again finding no evidence 
for large-scale IC emission at the level expected from the previously claimed non-thermal detections. This 
latter result holds for all physically
reasonable spatial distributions, with the most probable IC distribution providing an upper limit of
$2.7 \times 10^{-12}$ \fxunits.

The Coma cluster is a test bed galaxy cluster target for the \nustar\ X-ray observatory \citep{Harrison.ea:13}.
\nustar\ is the first focusing hard X-ray telescope in orbit, with the ability to focus X-rays above 10 keV. 
\nustar\ operates in the wide energy band from 3 to 79 keV, carrying 
two identical co-aligned X-ray telescopes with an angular
resolution of 18\arcsec\ (FWHM). The focal planes of
each telescope, referred to as focal plane modules A and B, provide a spectral
resolution of 400 eV (FWHM) at 10 keV and a combined effective area at 30 keV of 220 cm$^2$.
The somewhat lower effective area compared to previous instruments is more than compensated for by the focusing 
capability which vastly reduces the background level and point source contamination. The 
$\sim 13$\arcmin $\times 13$\arcmin\ FOV is considerably smaller than collimators on board \rxte, \sax, and \suzaku\ 
which have quite large, $\gtrsim 1^{\circ}$ FOVs. 

We describe the \nustar\ Coma observation and its processing in 
Section 2. We show images in different energy bands in Section 3 and the spectral analysis, both for the global spectrum
and spatially resolved spectroscopy in the form of a temperature map in Section 4. The results are discussed in 
Section 5. The cosmology adopted in this paper assumes a flat universe with \ho = 70 \kmsmpc,
\omegam = 0.27 and \omegalambda = 0.73. All errors are quoted at the 68\% confidence
limit. At the redshift of Coma, $z = 0.0231$,  1\arcmin\ subtends 28 kpc.

\begin{figure}[tdh]
\parbox{0.5\textwidth}{
\includegraphics[width=0.5\textwidth]{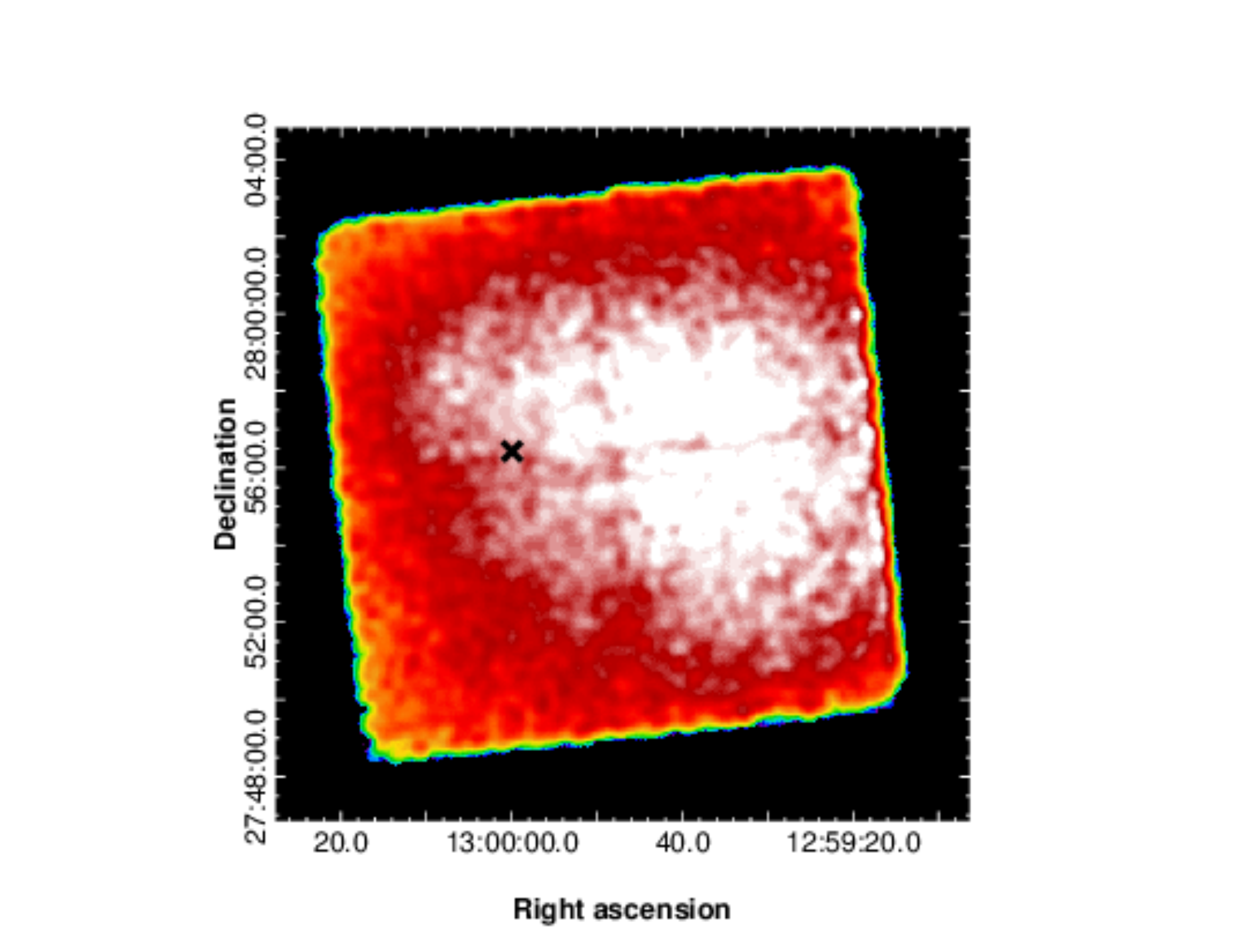}}
\parbox{0.5\textwidth}{
\includegraphics[width=0.5\textwidth]{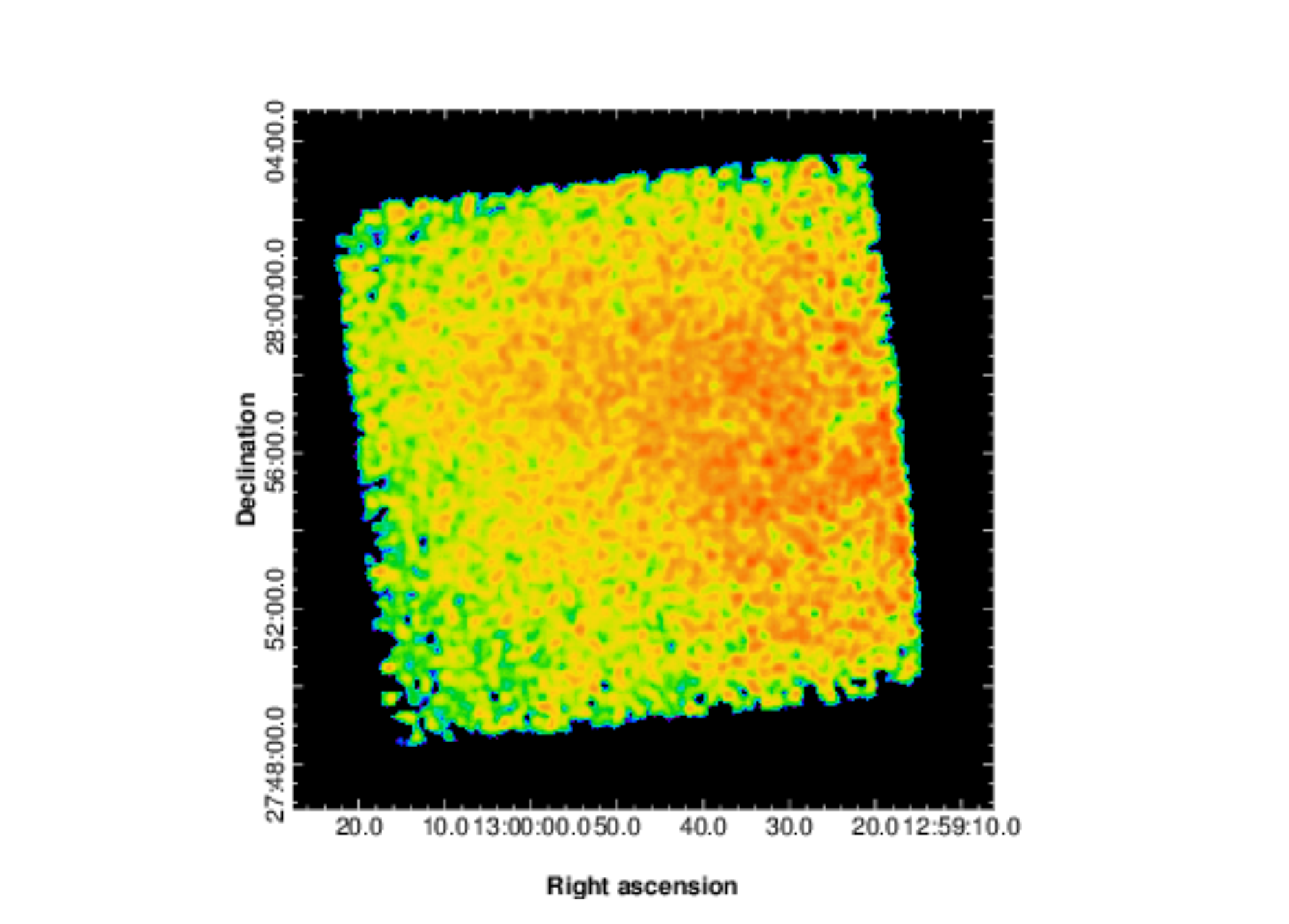}}
\parbox{0.5\textwidth}{
\includegraphics[width=0.5\textwidth]{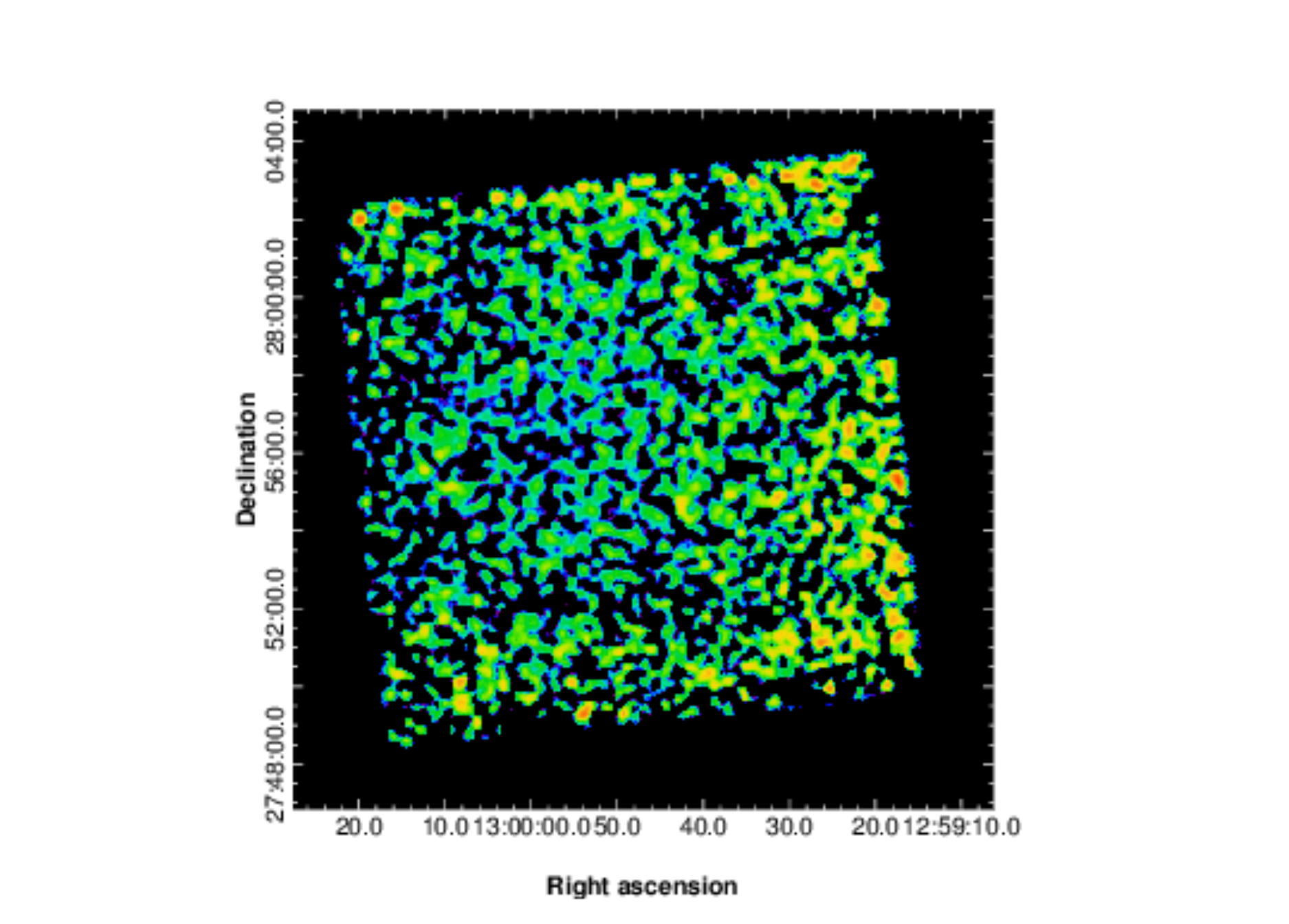}}
\caption{\label{images} \footnotesize
\nustar\ background-subtracted and exposure-corrected images of Coma, combining both telescopes A and B. Images are presented on a common logarithmic scale from 
0 counts pixel$^{-1}$ to $1.1\times10^{-4}$ pixel$^{-1}$. The energy band for each image is from top to bottom: 3--10 keV, 10--20 keV and 20--50 keV.
Images have been smoothed with a Gaussian kernel of 17$^{\prime\prime}$.2 (7 pixels). The x point in the 3-10 keV image marks the average position of the optical axis.} 
\end{figure}

\section{Observation and data processing}
\label{obs}
The Coma Cluster was observed by \nustar\ 
on 2013 June 16-17 for a total unfiltered exposure time
of 115 ks. 
We processed the data from both modules A and B using standard pipeline processing (HEAsoft v6.15.1 and NuSTARDAS v1.3.1)
and the 20131223 version of the \nustar\ Calibration Database. 
The data were filtered for periods of Earth occultation, high instrumental background
due to South Atlantic Anomaly (SAA) passages, and known bad/noisy detector pixels.
We adopted strict criteria regarding passages through the SAA and a tentacle-like region of higher activity near part of the SAA; 
in the call to the general processing routine that creates Level 2 data products, {\ttfamily nupipeline}, the following flags were included:
SAAMODE=STRICT and TENTACLE=yes. The resulting clean exposure time is 55 ks.
The level of solar activity during the observation was at the B level (the X-ray flux
level as registered by the \emph{GOES} satellite was below $10^{-6}$ W m$^{-2}$ in the
1-8 \AA\ wavelength range), not sufficient to produce any significant reflected solar stray light 
\citep{Wik.ea:14}. The lack of solar activity is also corroborated by the absence of variability
in light curves extracted from the cleaned event files.

From the cleaned event files, we extracted images and light curves using
{\ttfamily xselect}, created exposure maps using {\ttfamily nuexpomap}, and extracted
spectra and associated response matrix (RMF) and
auxiliary response (ARF) files using {\ttfamily nuproducts}. The
call to {\ttfamily nuproducts} included {\ttfamily extended=yes}, most appropriate
for extended sources, which weights the RMF and
ARF based on the distribution of events within the extraction
region, assuming that to be equivalent to the
true extent of the source. The effective smoothing
of the source due to the point spread function (PSF)
is not taken into account and it will be implemented in future analysis
of extended sources. However given the relatively narrow
FWHM of 18\arcsec\ this omission is not impacting the analysis
given the angular size of the regions considered for spectral fitting in this work.

The \nustar\ mirrors have a design based on the Wolter-I approximation, which is a 
double mirror design that focuses X-rays with two grazing angle reflections. It is possible for 
photons at very shallow or very steep angles to be reflected only once by the
mirror assembly. These ``ghost rays'' can originate from sources between 3\arcmin\ and 40\arcmin\
off-axis. Based on the \xmm\ pn Coma mosaic image of \citet{Wik.ea:09} we simulated by ray-tracing
the impact of ghost rays in the \nustar\ observation. They are affecting mainly the regions at large
off-axis angles from the optical axis, shown in the top panel of Figure \ref{images} and obtained as the peak of the exposure map, given
the motion of the optical axis due to the thermal expansion of the mast. Ghost rays are contributing at most 18\% on average of the total emission
in the western regions of the observation. The ray-tracing simulation shows that there might a be bias on the measured temperatures which is below 3\% for ghost ray fractions 
below 20\% and below 8\% for a ghost ray fraction between 20\% and 30\%. A preliminary comparison with \xmm\ and \chandra\ data shows an increasing higher flux in \nustar\ 
with increasing distance from the optical axis at a level consistent with the expected ghost-ray  contamination but no systematic discrepancy in the temperature determination.


\section{Image analysis}

We exploit \nustar's unprecedented focusing capabilities and spatial resolution to create images of the Coma Cluster 
in various energy bands: 3--10 keV and, for the first time, in the high energy bands 10--20 keV and 20--50 keV.

We perform exposure-correction using the task {\ttfamily nuexpomap}, creating exposure maps at single energies
for each band, roughly corresponding to the mean emission-weighted energy of the band.
Background images were produced using {\ttfamily nuskybgd} \citep{Wik.ea:14}. The images have been Gaussian smoothed
by 17$^{\prime\prime}$.2 (7 pixels) to be consistent with the PSF's FWHM of $\sim 18$\arcsec. 
Background-subtracted and exposure-corrected images in the three energy bands for the co-added instruments A and B are presented in Figure \ref{images}.
The 3--10 keV and 10--20 keV images show the well known morphology of the hot gas in the center of the Coma Cluster as shown by
\rosat, \xmm, and \chandra, whereas the 20--50 keV band, though background dominated, shows an excess to the west where hotter temperatures are found (see Section 4.2).

\begin{figure}[th]
\hspace{-0.5truecm}
\includegraphics[width=0.35\textwidth,angle=-90]{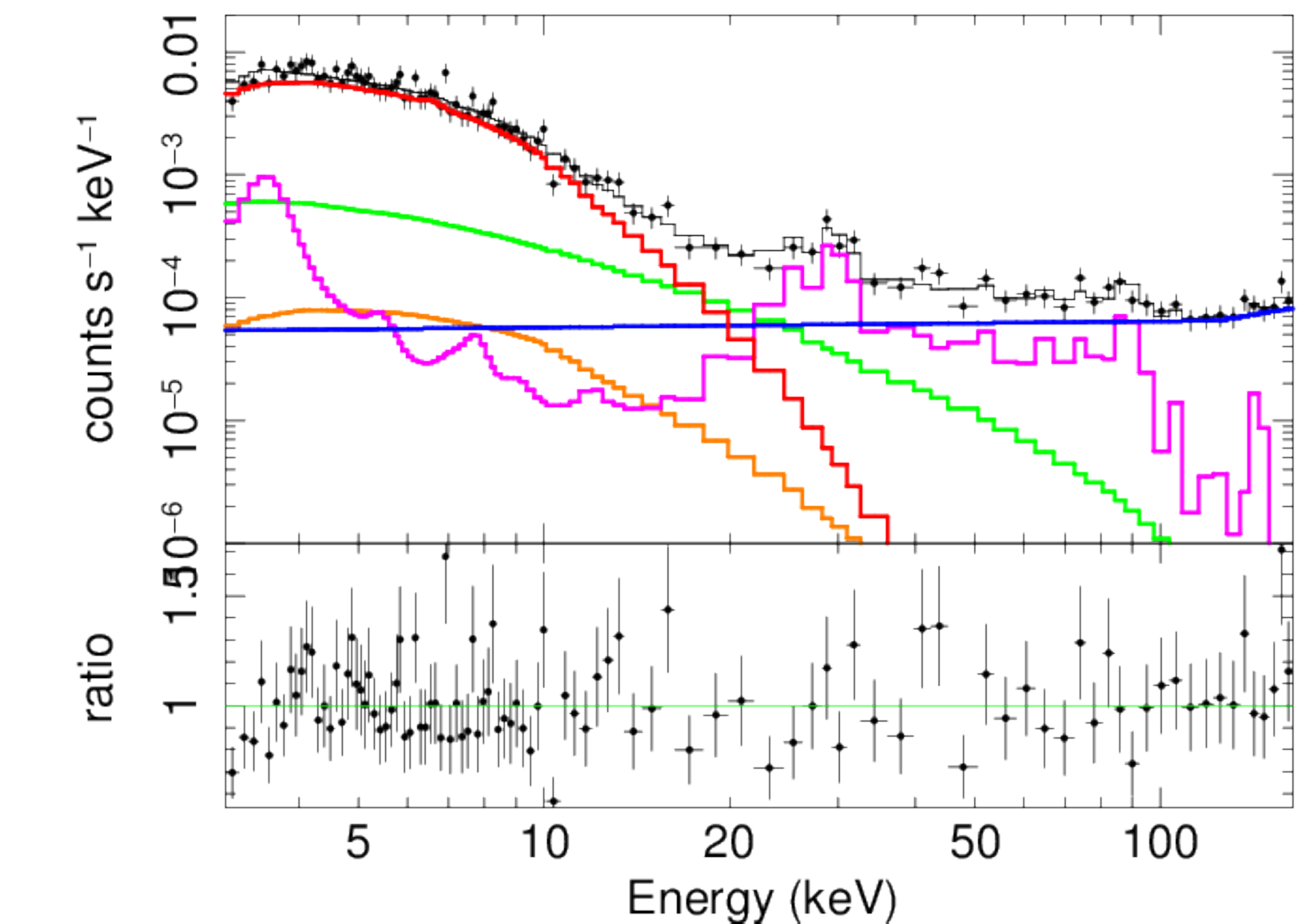}
\caption{\label{examplefit} Detector B spectrum of one of the regions (region 4) shown in Figure \ref{mapregions} extracted to obtain the temperature map discussed in 
Section \ref{sectiontmap}. The various background components have been modeled (blue: instrumental particle background continuum; magenta: instrumental lines and solar reflected component; orange: focused cosmic X-ray background (FCXB); green: aperture background) and the source component is shown by
the solid red line. The ratio of data over the model are also shown.\footnotesize}
\end{figure}

\section{Spectral analysis}
\label{spectral}

In Section 4.1 we examine the global spectrum observed by \nustar, and in Section 4.2 we break that spectrum up into a spatially-resolved 6$\times$6 grid of spectra to probe 
temperature variations. We can also constrain the relative calibration between the two \nustar\ telescopes.

We fitted the spectra of two detectors A and B separately with an \apec\ thermal plasma \citep{Smith.ea:01} modified by Galactic
absorption \citep{Kalberla.ea:05} fixed at $N_{\rm H} = 8.58 \times 10^{19}$ cm$^{2}$. This absorption has negligible effect in the \nustar\
bandpass. 
The spectral fitting was performed with \xspec\ \citep{Arnaud:96} in the 3-120 keV band using the $C$-statistic and quoted
metallicities are relative to the abundances of \citet{Anders.ea:89}. Energies in the range 79 keV $< E <$ 120 keV are primarily
used to constrain the instrumental background (in a similar fashion to the use of energies above 10 keV for satellites operating in the 0.5-10 keV energy band
such as \xmm) and their inclusion does not significantly contribute to the $C$-statistic or the 
resulting best-fit parameters. 
Although not strictly necessary for a fit using the $C$-statistic, we re-binned the data to ensure a minimum 30 counts per bin, reducing the time
required to perform fits and emphasizing differences between the model and the data.

To account for the background we included the spectral components of the \nustar\ background described
in detail in \citet{Wik.ea:14}. They can be characterized as originating from (1) instrument Compton scattered continuum emission; (2) instrument activation and 
emission lines; (3) cosmic X-ray background from the sky leaking past the aperture stops (Aperture); (4) reflected solar X-rays (Solar); (5) focused and ghost ray cosmic X-ray background (FCXB).
We do not have regions free of cluster emission so we can not apply straightforwardly the procedure
adopted by {\ttfamily nuskybgd}. We have an empirical nominal model based on blank field observations that we adopted in the
fit of the various regions (see Figure \ref{examplefit} for an example). 

In order to determine the best fit value and confidence interval for the spectral parameter of interests
we used Bayesian statistics and a Markov chain Monte Carlo (MCMC) technique. We performed MCMC simulations
using the \xspec\ implementation of the algorithm of \citet{Goodman.ea:10} where an ensemble of "walkers", which
are vectors of the fit parameters, are evolved via random steps determined by the difference between two walkers.
We evolved eight walkers for a total of $10^4$ steps, after discarding the initial 5000 steps (``burn-in'' phase) to ensure the chain
reached a steady state.
We turned on the Bayesian statistic setting up Gaussian priors centered on the expected value forecasted by
{\ttfamily nuskybgd} for the particular region of interest. We set widths equal to the expected systematic error for the various background 
normalizations (8\% for the aperture component, 3\% for the instrumental continuum, 50\% on the FCXB, 10\% on 
the solar component), and used constant priors for the temperature, abundance and normalization of the 
\apec\ cluster thermal component.
We then marginalized over all the other parameters to generate posterior probabilities for the parameter of interest, such
as the temperature or the normalization of the thermal component, using the \xspec\ command {\ttfamily margin}. The results found with 
this method are consistent with the procedure adopted in \citet{Wik.ea:14}. For ease of presentation we will show in the following 
figures the background-subtracted spectra using the realizations provided by {\ttfamily nuskybgd}.

\subsection{Global Spectrum}

\begin{figure}[th]
\hspace{-0.5truecm}
\includegraphics[width=0.35\textwidth,angle=-90]{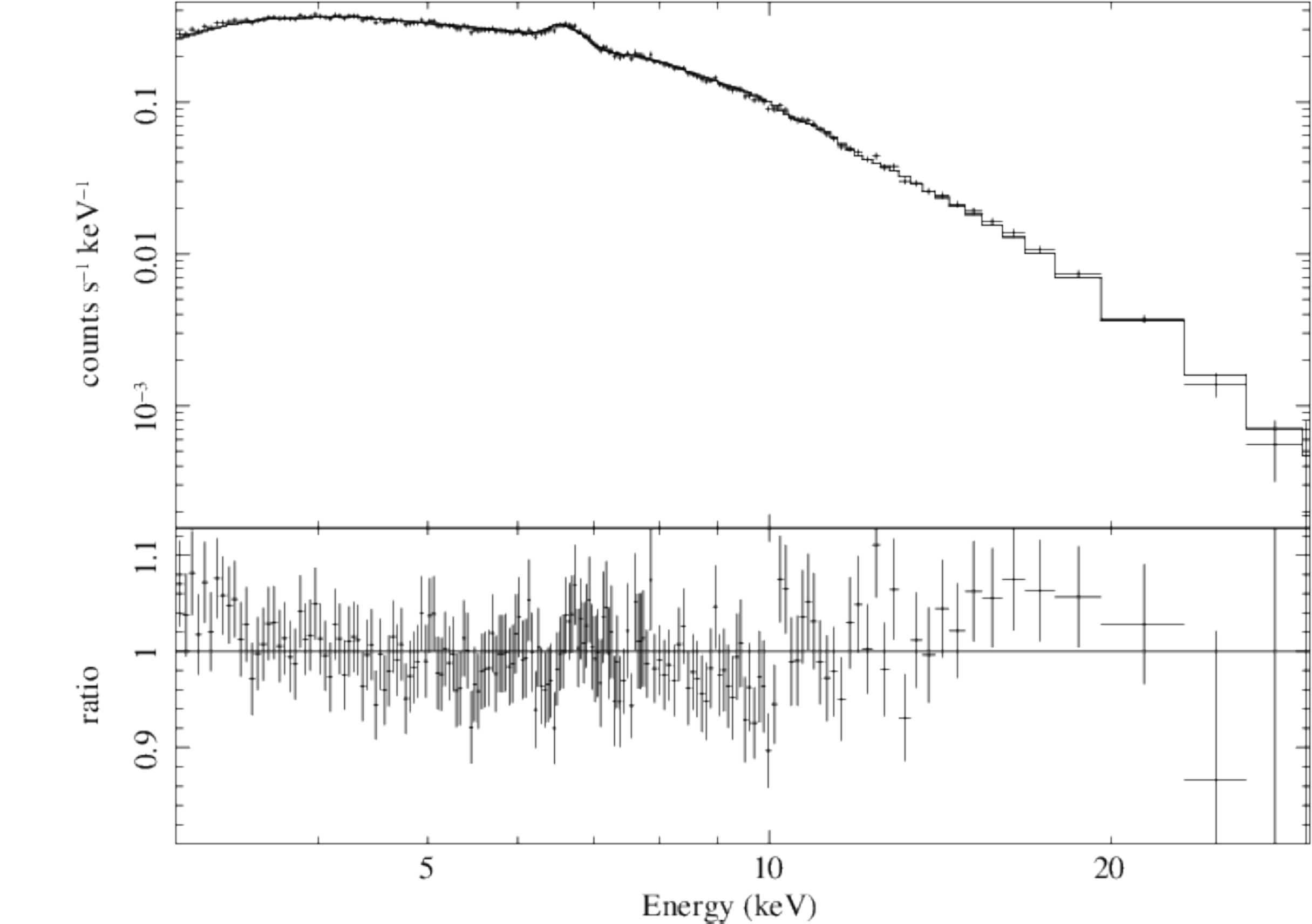}
\caption{\label{total1T} Background-subtracted Coma Cluster spectrum extracted from the central 12\arcmin\ $\times$ 12\arcmin\ region. The spectra of detector A and B have been combined for 
clarity. The best-fit 1T model and ratio of the spectrum over the model are also shown.\footnotesize}
\end{figure}

\begin{deluxetable*}{lcccccc}
\tablewidth{0pt}
\tablecaption{Global Spectrum Fit Parameters
\label{tab:fit}}
\tablehead{
    & $kT$ 
    & abund.\   
    & Norm.\tablenotemark{a} 
    & $kT$ or $\Gamma$ 
    & Norm.\tablenotemark{a} or IC flux\tablenotemark{b} 
    & $C$-stat/dof \\
        & & & & & ($10^{-1}$ cm$^{-5}$ or  & \\
 Model (Energy Band (keV)) & (keV) 
            & (\zsun) 
            & ($10^{-1}$ cm$^{-5}$)    
            & (keV or $\cdots$)  
            & $10^{-12}$ ergs s$^{-1}$ cm$^{-2}$)
            & 
}
\startdata
1T 3--120\tablenotemark{c}    &    $8.52^{+0.03}_{-0.04}$  
         &    $0.178^{+0.006}_{-0.004}$ 
         &    $1.042^{+0.004}_{-0.005}$  
         &    $\cdots$     
         &     $\cdots$
         &     $3196\pm2$\tablenotemark{d}/2975\\
1T 3--30    &    $8.58^{+0.10}_{-0.04}$    
         &    $0.178 \pm 0.007$
         &    $1.044^{+0.006}_{-0.008}$
         &   $\cdots$  
         &    $\cdots$ 
         &    1216/1127 \\
1T 3--10    &    $8.20\pm0.08$    
         &    $0.179 \pm 0.007$
         &    $1.068\pm0.007$
         &   $\cdots$  
         &     $\cdots$
         &    356/344 \\
1T 5--10     &    $8.54\pm0.12$    
         &    $0.19 \pm 0.01$
         &    $1.026^{+0.015}_{-0.013}$
         &   $\cdots$  
         &     $\cdots$
         &    250/244 \\
1T 5--30     &    $8.90\pm0.07$    
         &    $0.20 \pm 0.01$
         &    $1.003\pm0.008$
         &   $\cdots$  
         &     $\cdots$
         &    1078/1027 \\
2T 3--30     &   $9.03\pm0.11$    
         &     $0.213 \pm 0.09$
         &    $0.988 \pm 0.013$
         &    $1.02\pm0.21$  
         &    $0.794^{+0.679}_{-0.275}$
         &    1151/1125 \\
2T 4--30     &   $9.02^{+0.11}_{-0.10}$    
         &     $0.214 \pm 0.010$
         &    $0.987\pm0.012$
         &    $0.95^{+0.52}_{-0.23}$  
         &    $0.794^{+0.679}_{-0.275}$
         &    1113/1075 \\
$T_{\rm{map}}$ 3--30\tablenotemark{e}  &  $\cdots$
         &   $\cdots$
         &     $\cdots$
         &     $\cdots$
         &     $\cdots$
         &    1208/1129 \\
$T_{\rm{map}}$ 4--30\tablenotemark{e}  &  $\cdots$
         &   $\cdots$
         &     $\cdots$
         &     $\cdots$
         &     $\cdots$
         &    1137/1079 \\
$T_{\rm{map}}$+IC 4--30  &   -
         &    $\cdots$
         &    $\cdots$
         &   2.0 (fixed)   
         &   $< 0.48$
         &   1137/1080 \\
$T_{\rm{map}}$+IC 4--30\tablenotemark{f} &   -
         &    $\cdots$
         &    $\cdots$
         &   2.0 (fixed)   
         &   $< 5.1$
         &   1130/1078
 \enddata
\tablenotetext{a}{Normalization of the APEC thermal spectrum,
which is given by $\{ 10^{-14} / [ 4 \pi (1+z)^2 D_A^2 ] \} \, \int n_e n_{\rm{H}}
\, dV$, where $z$ is the redshift, $D_A$ is the angular diameter distance,
$n_e$ is the electron density, $n_{\rm{H}}$ is the ionized hydrogen density,
and $V$ is the volume of the cluster.}
\tablenotetext{b}{20--80~keV.}
\tablenotetext{c}{Obtained with the Bayesian MCMC analysis.}
\tablenotetext{d}{Mean and standard deviation of the fit statistic values over the steps of the chain.}
\tablenotetext{f}{For these fits we quote only the resulting fit statistic value. The only fitting parameters are the
overall normalization of the sum of the 36 thermal models obtained from the regions used for the temperature map.}
\tablenotetext{f}{This fit has been obtained by thawing the overall normalization constants of the $T_{\rm{map}}$
for the two detectors A and B, see text for details.}
\end{deluxetable*}

To compare with results obtained with other satellites we extracted the global A plus B spectra from a box of
12\arcmin$\times$12\arcmin\ encompassing 85\% of the  FOV. The spectra have excellent statistical quality with 
$\sim 2.5 \times 10^{5}$ total counts with a source contribution to the total emission of 89\% in the 3-30 keV energy band.

We first consider a single temperature (1T) model fit to the data, which is the simplest possible description of the spectrum.
This is unlikely to be a realistic description as it is known that even the very center of Coma hosts temperature
variations \citep[e.g.,][]{Sanders.ea:13b}. However multi-temperature, featureless, spectra with a range of temperatures can be 
well fitted by a 1T model \citep[e.g.,][]{Mazzotta.ea:04}. We find a temperature of $8.52$ keV as a peak of the marginalized 
posterior distribution, with a 68\% confidence interval of [8.48,8.55] keV. As shown in Table \ref{tab:fit},
a fit in the 3--30 keV energy band obtained with background subtraction of a realization of the background model \citep[the procedure used in][]{Wik.ea:14} 
returns consistent results. In Figure \ref{total1T} we show the co-added A and B background-subtracted spectrum in the 3--30 keV energy band obtained with this latter method.
The spectrum is well, but not perfectly, described by an isothermal spectrum over an order of magnitude in energy.
We use the background-subtraction method to quickly explore the dependence of the temperature determination when using different energy bands for the 
spectral fitting. 
In the absence of systematic calibration issues, different temperatures returned when fitting different energy bands is 
yet another indication of a multi-temperature component spectrum. 
This is indeed the case for the Coma global spectrum as increasing either the upper end or the lower end of the baseline energy band 
increases the derived temperature, as detailed in Table \ref{tab:fit}.

The next step to add complexity to the fitting model is a two temperature (2T) model consisting of two APEC components with abundances tied
together. This model is routinely used when dealing with multi-temperature component spectra.
The fits improve, though the temperature found for the low-$T$ component ($1.02\pm0.21$ keV) does not represent any real temperature
in the spectrum. This seems more a result of the fit procedure that is accommodating the curvature of the residuals that are not well fitted
by a 1T model in the low energy part of the spectrum where the statistical quality of the data is higher. To support this hypothesis
we performed simulations with the \nustar\ responses of a two thermal component model with temperatures of 7 and 9 keV respectively.
We chose the ratio of the normalization of the two components to be equal to that which best approximates the observed spectrum. When a 2T model is applied, the 
fitting process favors a high--$T$ component of the order of 8--9 keV accounting for most of the emission in the fitted band and a lower--$T$ component
(0.5--1 keV), which improves the fit at the lower range of the energy band. Similar results when fitting a 2T model have been obtained by 
\citet{Ajello.ea:09} when fitting \xmm\ and \swift\ BAT data ($kT_{\rm{high}} = 8.40^{+0.25}_{-0.24}$ keV and $kT_{\rm{low}} = 1.45^{+0.21}_{-0.11}$ keV).
The low--$T$ component has been interpreted as due to thermal X-ray emission from the galaxies in Coma \citep[][]{Finoguenov.ea:04,Sun.ea:07}. 
While this might be a possible interpretation for satellites sensitive to energies down to 0.5 keV, it can be ruled out for emission above 3 keV as 
seen by \nustar.

Following the success of the \xmm-derived temperature map for explaining the thermal origin of the \suzaku\ HXD-PIN and \swift\ BAT high-energy
spectra \citep{Wik.ea:09,Wik.ea:11} we adopted the same approach exploiting the temperature map obtained by \nustar\ itself (discussed in the following 
Section \ref{sectiontmap}). 
We summed the 36 1T APEC models with temperature, abundances
and normalization fixed to construct a $T_{\rm{map}}$ model for which only the overall normalization was allowed to vary (an adjustment at the 2\% level)
to give a fit with the same quality as the 2T fit (see Table \ref{tab:fit}). The comparison in the 3-30 keV energy band (cstat/dof = 1208/1129 for the 
$T_{\rm{map}}$ model and 1151/1125 for the 2T model) points again to a slightly better fit for the 2T model, following the same lines discussed in the
above paragraph.

Armed with a reasonable model description of the multiple thermal components in the center of Coma as obtained by the $T_{\rm{map}}$ model, we proceed
to constrain the non-thermal flux in the \nustar\ spectrum. We fixed the photon spectral index of the non-thermal component to $\Gamma = 2$ based 
on previous analysis with \sax\ and \rxte\ \citep{Fusco-Femiano.ea:04,Rephaeli.ea:02} in order to have a direct comparison with those previous works;
if allowed to vary its value is unconstrained with a negative best fit value. With the best fit 
$T_{\rm{map}}$ model, the 90\% upper limit on the 20--80 keV non-thermal flux is $4.8 \times 10^{-13}$ \fxunits; if we allow the overall normalization constant of 
the  $T_{\rm{map}}$ model to vary we obtain a best fit value of $0.99\pm0.01$, and the 90\% upper limit on the 20--80 keV non-thermal flux becomes $5.1 \times 10^{-12}$ \fxunits.

We investigated a possible bias arising by a distributed non-thermal component which could
bias upward the temperatures in the temperature map. We therefore re-measure normalizations and temperatures of the thermal 
components just by fitting the 1T models in the 3--10 keV energy band in order to minimize the impact of the eventual presence
of the non-thermal component. The temperatures thus obtained are on average 5\% lower and the normalizations 2\% higher. With this $T_{\rm{map}}$ model
we then constrained the non-thermal flux as done above. The 90\% upper limit on the 20--80 keV non-thermal flux is 
$1.7 \times 10^{-12}$ \fxunits; if we allow the overall normalization constant of the  $T_{\rm{map}}$ model to vary down to the value of 0.99, the 90\% upper limit on the 20-80 keV non-thermal flux 
becomes $4.7 \times 10^{-12}$ \fxunits which is 8\% smaller than the value derive with the $T_{\rm{map}}$ model obtained using the full energy band.
We therefore conclude that the 3--10 keV  $T_{\rm{map}}$ model did not result in a significant higher upper limit on the IC flux.

We remark that the results from the two detectors are in very good agreement when fitted individually against the same model. When fitted jointly
the constant introduced in the models to allow for different normalizations of the spectral components is of the order $0.998\pm0.004$.

\subsection{Temperature Map}
\label{sectiontmap}

\begin{figure}[t]
\centerline{
\includegraphics[width=0.6\textwidth]{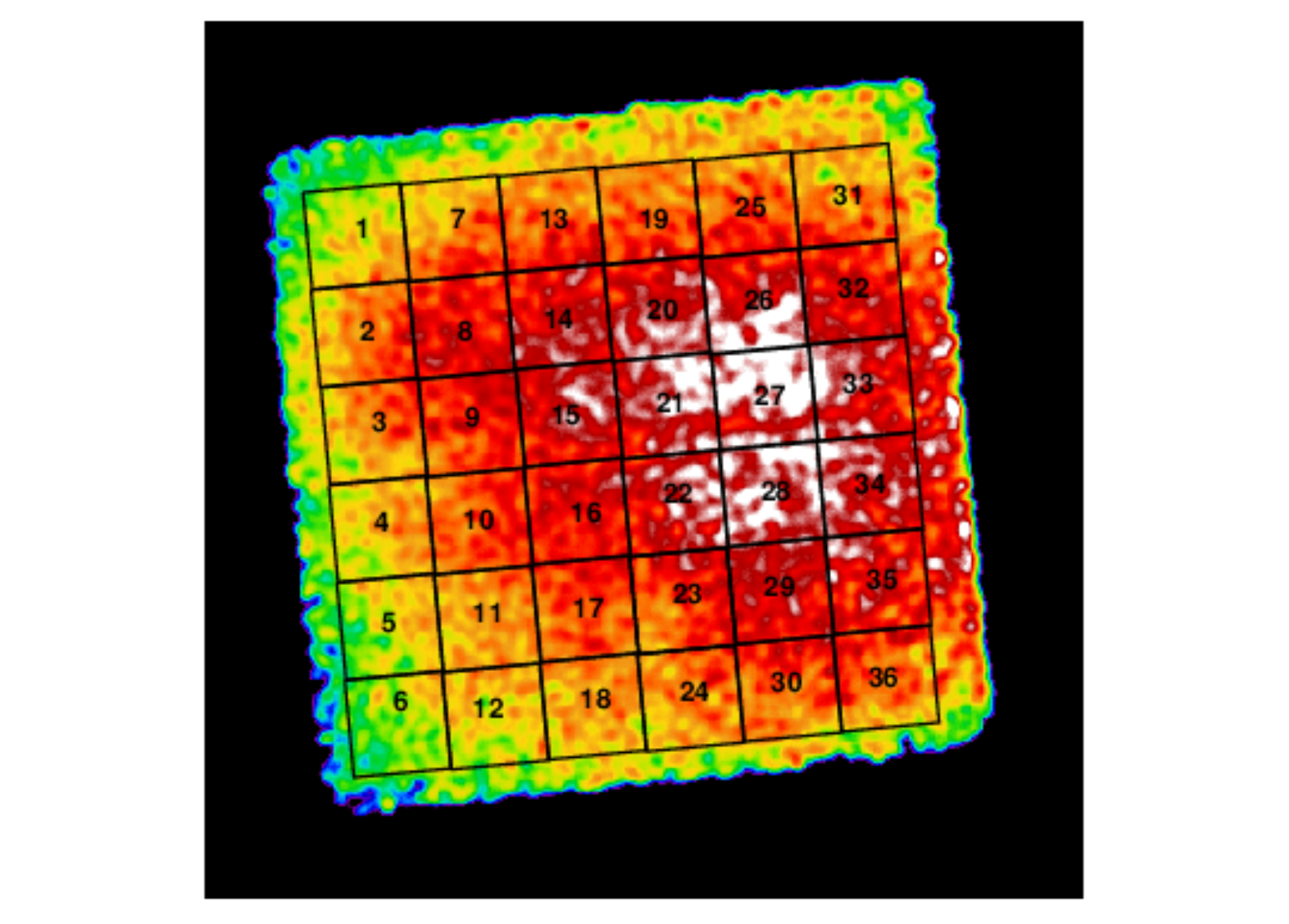}
}
\caption{\label{mapregions}Superimposed on the \nustar\ 3--10 keV image of the Coma Cluster, we plot
the regions used for spatially resolved spectral extractions. The spectrum of each region is fitted with a single temperature model 
to build the temperature maps shown in Figure \ref{tmap}.\footnotesize}
\end{figure}

\begin{figure*}[tdh]
\vspace{-0.5cm}
\centerline{
\parbox{0.5\textwidth}{
\includegraphics[width=0.6\textwidth]{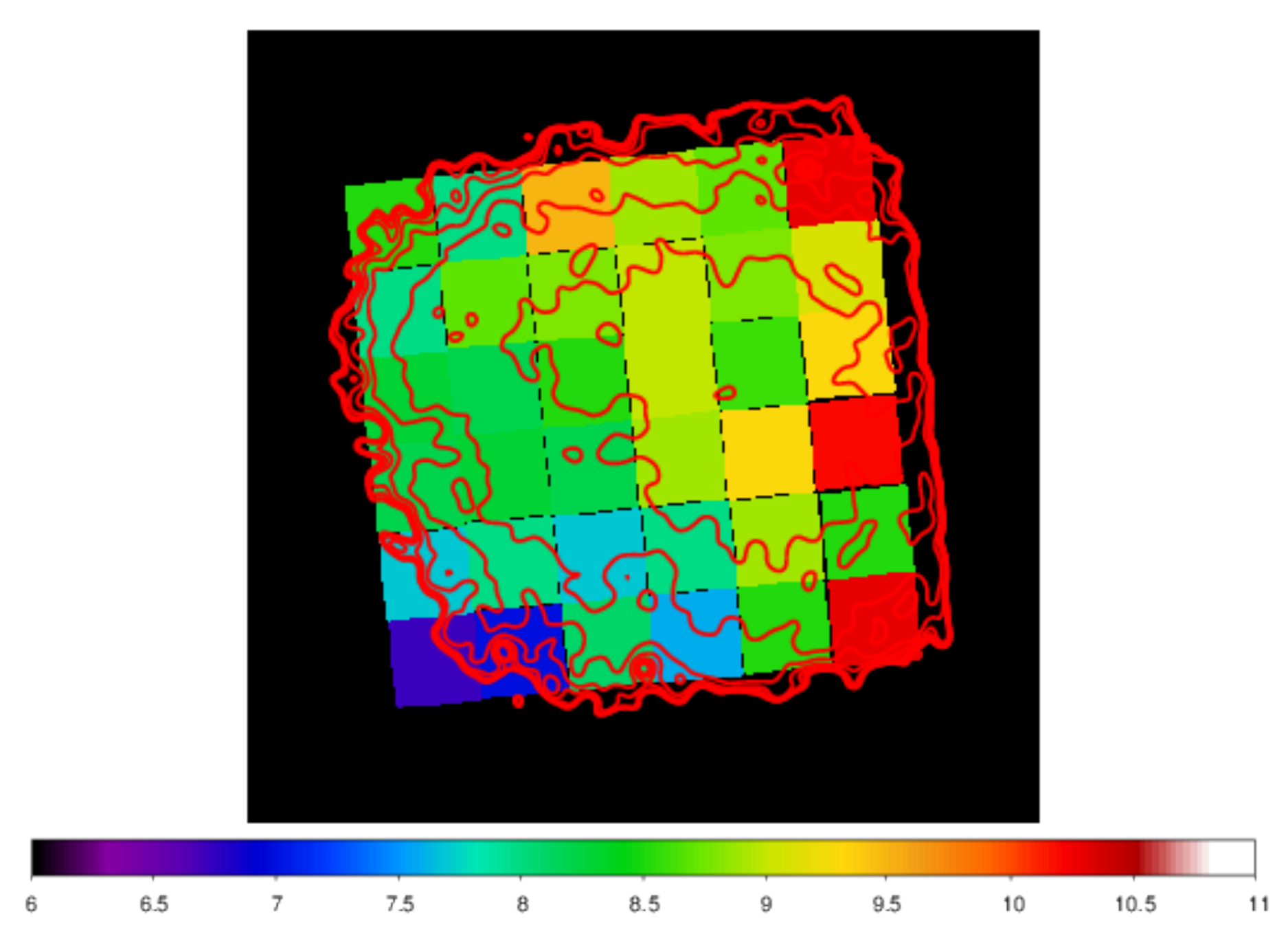}}
\parbox{0.5\textwidth}{
\includegraphics[width=0.5\textwidth]{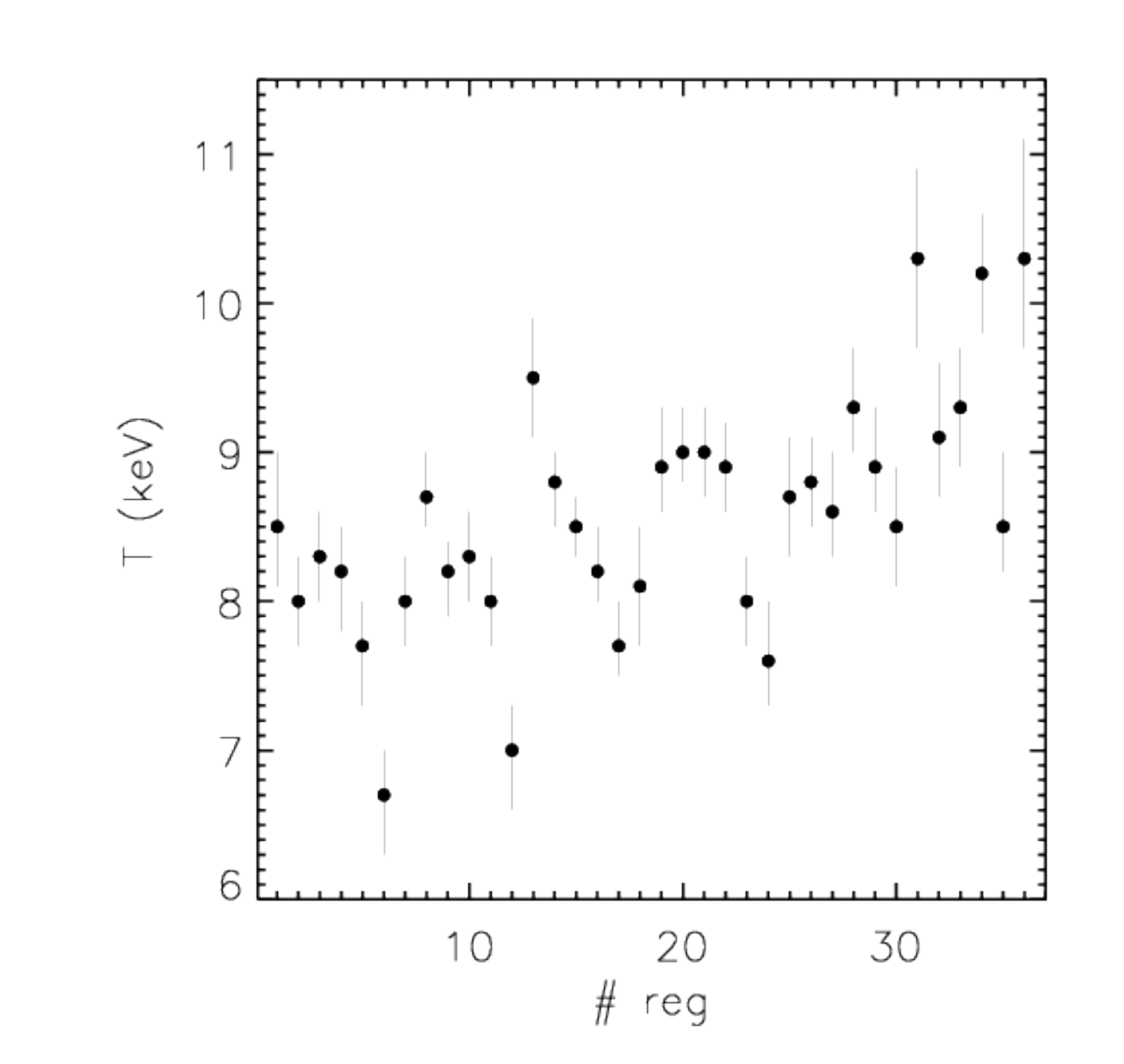}}
}
\caption{\label{tmap} \footnotesize
Left panel: temperature map of the central region of Coma obtained from a joint fit of the detectors A and B spectra extracted from the 
regions depicted in Figure \ref{mapregions}. Overlaid are the surface brightness contours obtained by the \nustar\ image in the 3--10 keV energy band.
Right panel: values and 1$\sigma$ error bars for the temperature in each region of the temperature map.} 
\end{figure*}

\begin{figure*}[tdh]
\vspace{-0.5cm}
\centerline{
\parbox{0.5\textwidth}{
\includegraphics[width=0.5\textwidth]{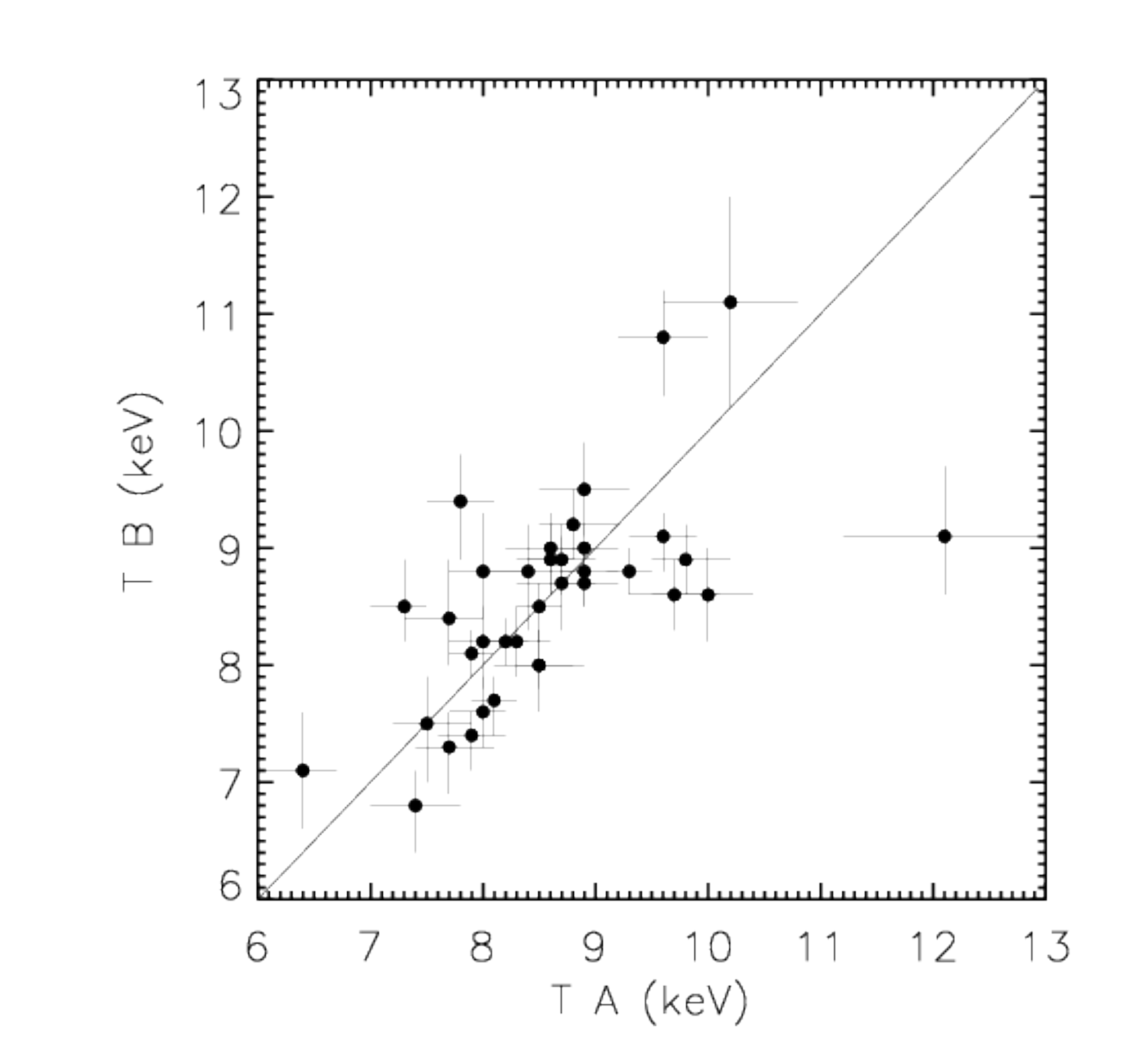}}
\parbox{0.5\textwidth}{
\includegraphics[width=0.5\textwidth]{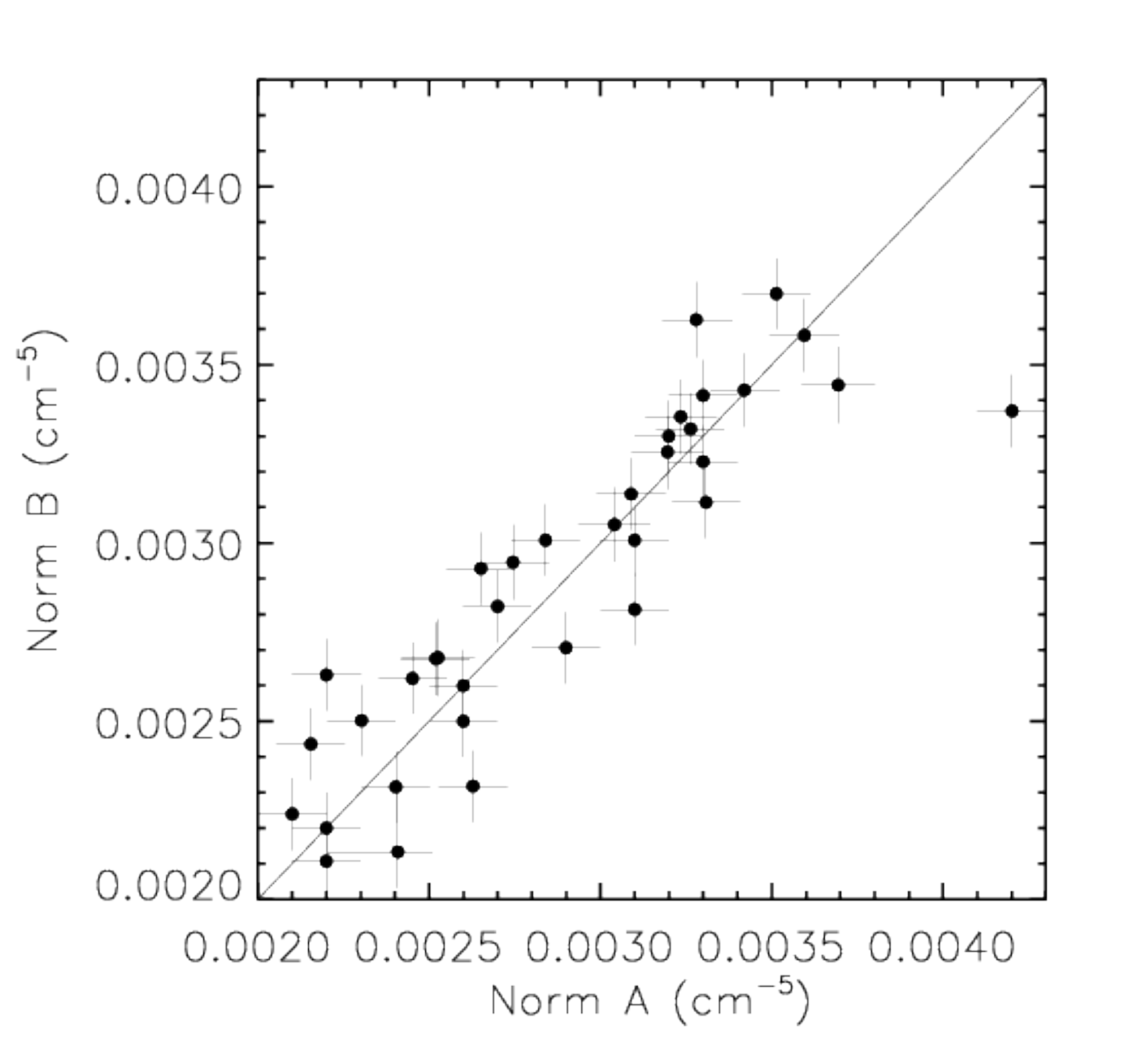}}
}
\caption{\label{comptmap} \footnotesize
Left panel: comparison of the temperature found by the two detectors A and B in the regions of the temperature map of Figure \ref{tmap}.
Right panel:  comparison of the normalization found by the two detectors A and B in the regions of the temperature map of Figure \ref{tmap}.} 
\end{figure*}

In order to study the cluster temperature structure, we extracted spectra
in 2\arcmin $\times$ 2\arcmin\ contiguous regions in the sky, as plotted in
Figure \ref{mapregions}.
With the same approach depicted in the previous paragraph for the global spectrum
we fitted 1T models for the cluster component plus the spectral components needed
to model the sky and instrumental background.

The temperature map thus obtained is shown in Figure \ref{tmap}. The overall trend
is a temperature gradient from the hotter northwest regions, with temperatures in the 9-10 keV
range, to the cooler regions in the southeast, with temperatures of the order 7 keV (regions 6 and 12
in Figure \ref{tmap}). The spectra are well fitted by 1T models. Models with additional spectral components 
(e.g. 2T) do not significantly improve the fit.

As an additional check of the cross calibration between the two detectors A and B, we compare the results for temperature and normalization
of the best fit thermal model for the regions of the map in Figure \ref{comptmap}. We again find good agreement between the two detectors:
the ratio of the temperatures found with A with respect to the temperatures found with B has a mean of 1.01 with a standard deviation of 0.10.
The ratio of the normalization has a mean of 0.99 with a standard deviation of 0.07.

\begin{figure*}[tdh]
\centerline{
\parbox{0.5\textwidth}{
\includegraphics[height=0.28\textheight]{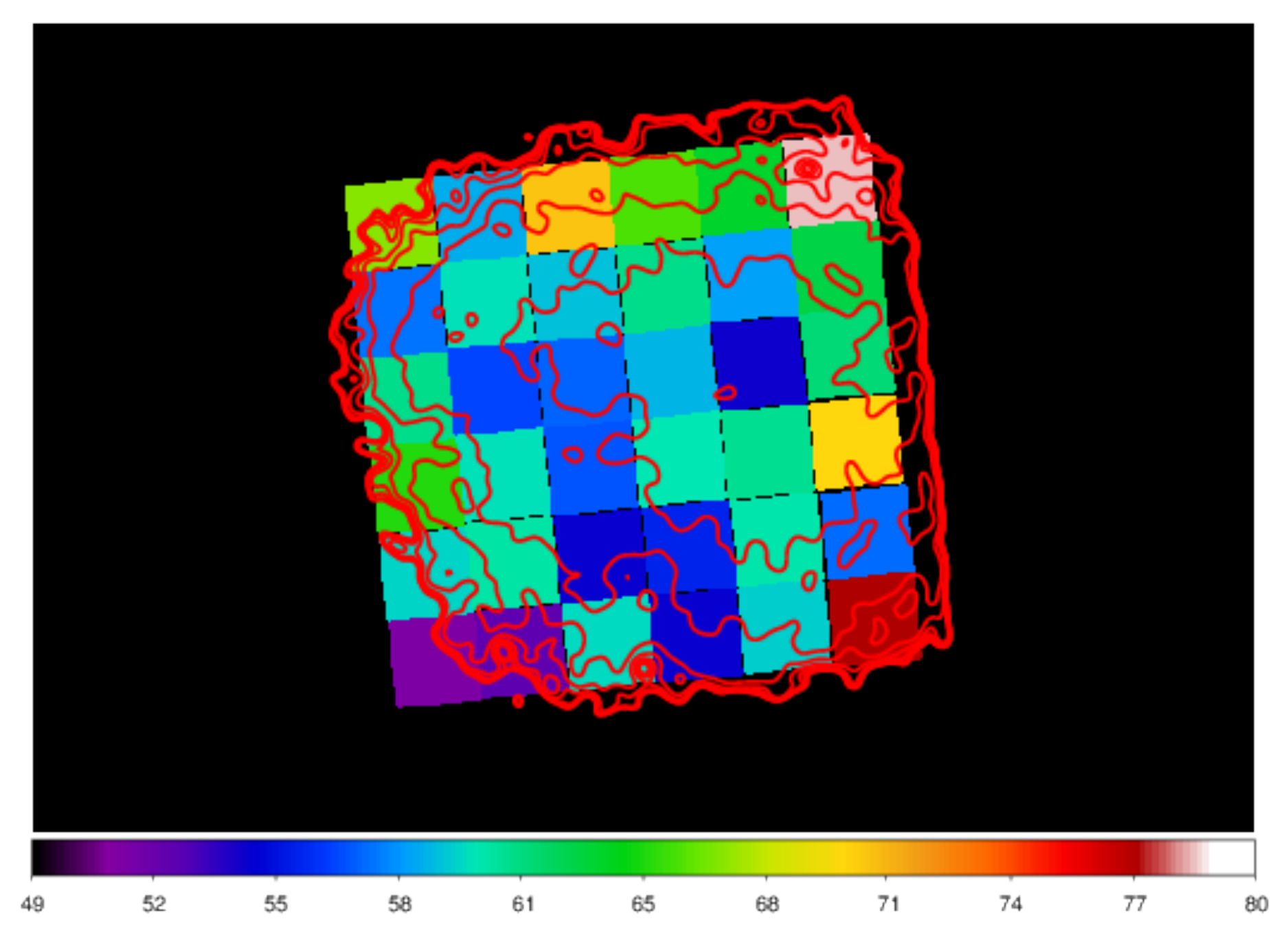}}
\parbox{0.5\textwidth}{
\includegraphics[height=0.28\textheight]{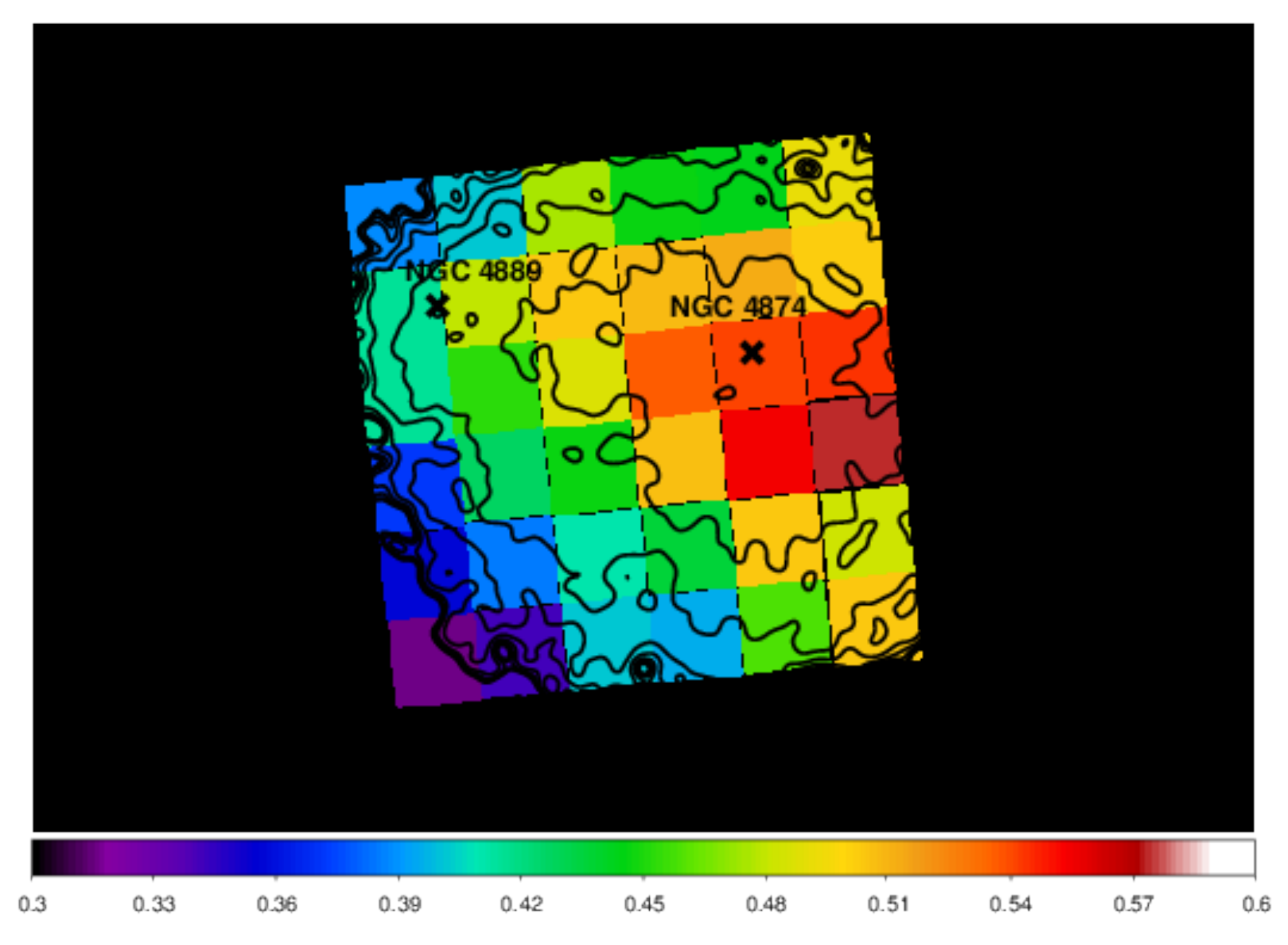}}
}
\caption{\label{psmaps} \footnotesize
Left panel: \nustar\ pseudo entropy map of the center of the Coma Cluster (arbitrary units). Overlayed are the surface brightness contours obtained by the \nustar\ image in the 3--10 keV energy band.
Right panel: \nustar\ pseudo pressure map of the center of the Coma Cluster (arbitrary units). Overlaid are the surface brightness contours obtained by the \nustar\ image in the 3--10 keV energy band and the positions of the brightest central galaxies, NGC 4889 and NGC 4874, are marked by x points.} 
\end{figure*}

\section{Discussion}

\subsection{The Global Spectrum and the IC Upper Limit}

The \nustar\ global spectrum extracted from a 12\arcmin $\times$ 12\arcmin\
region is not perfectly isothermal, as shown by both the fact that the measured temperature 
depends on the fitting energy range (Table \ref{tab:fit}) and by the temperature map presented in Figure \ref{tmap}. We therefore adopt a multi-temperature model
based on the temperature map in order to impose more accurate constraints on the non-thermal flux.
Depending on 2\% differences in the normalization of the multi-temperature model, the 90\%
upper limit to the IC non-thermal emission is in the range $4.8 \times 10^{-13}$ - $5.1 \times 10^{-12}$ 
\fxunits\ in the 20-80 keV band. The latter limit is no more stringent than the most recent limits.
In particular, we note that the region analyzed corresponds to only a small fraction of the extension of the non-thermal 
component. If for example we adopt disks of uniform surface brightness with various extension as in \citet{Wik.ea:11},
we see that the \nustar\ pointing can in principle cover from 7\% down to 1\% of the emission in the case of a disk of radius
25\arcmin\ and 60\arcmin, respectively. Clearly this \nustar\ pointing at a restricted central region of Coma where the thermal emission
is strongest is not optimal for investigating the presence of an IC component. A venue to address this problem in the future will be 
a \nustar\ mosaic covering the Coma Cluster.

\subsection{The Coma Center Thermodynamical Maps}

There is good agreement between the \chandra\ temperature map presented in \citet{Sanders.ea:13b}
and the \nustar\ map shown in Section \ref{sectiontmap}. 
The gradient in the emission-weighted projected temperature from the hot northwest side of the core to the cooler
southeast as seen by \chandra\ is nicely confirmed by \nustar, to a better level than the more uniform temperature
distribution seen by \xmm\ \citep{Arnaud.ea:01}. In fact, the ``hot spot'' seen in the \asca\ temperature map by \citet{Donnelly.ea:99}
can be understood in the context of this gradient, although not at the overall level of 13 keV suggested by the \asca\ temperature map.
The southeast cooler emission is a feature revealed at larger scales than the current \nustar\ observation 
by many satellites. It is related to a filament pointing toward NGC 4911 and NGC 4921, first discovered in \rosat\ observations \citep{vikhlinin.ea:99};
in the regions probed by the \nustar\ observation, the cooler emission is connected to the 
excess brightness linear features seen in the deep \chandra\ exposure of \citet{Sanders.ea:13b}. 

The temperature map of the Coma center together with the Bullet observation \citep{Wik.ea:14} 
marks the beginning of spatially resolved spectroscopy in the hard X-rays, in the energy band above 10 keV for galaxy clusters. The combination 
of relatively flat effective area in the 3-15 keV energy band (see Figure 5 of the \nustar\ Observatory Guide \footnote[1]{https://heasarc.gsfc.nasa.gov/docs/nustar/NuSTAR\_observatory\_guide-v1.0.pdf}) and relatively low background makes \nustar\ an ideal instrument
for measuring hot temperatures in galaxy clusters, and a benchmark comparison for the temperature measurement of more traditional satellites
in the 0.5-10 keV energy bands.  These are affected by steeply falling effective areas and flat or increasing background at high energies.
It will also be interesting to investigate the bias of the spectroscopic-like temperatures \citep[e.g.,][]{Mazzotta.ea:04} with the broader energy band 
of \nustar.

Using the projected temperature, $T$, and normalization, EM, values we can derive 
the projected entropy $T/\rm{EM}^{1/3}$ and the projected pressure $T \times \rm{EM}^{1/2}$.
This can be used to quickly explore relevant features in the intracluster medium \citep[e.g.,][]{Rossetti.ea:07}.
They are shown in Figure \ref{psmaps} and they highlight the disturbed state of the center of the Coma Cluster.
The entropy map is rather uniform on the scale the \nustar\ pointing ($\sim 200$ kpc) 
and this is typically observed in the center of mergers, non cool-core systems \citep[e.g.,][]{Cavagnolo.ea:09}.
The pressure map follows more closely the surface brightness distribution and its elongation in the axis connecting the
two central galaxies, NGC 4889 and NGC 4874. This is consistent with the picture of the presence of sub-halos associated
with the two galaxies perturbing the gravitational potential of the central region of the cluster \citep{Andrade-Santos.ea:13}.

\section{Summary 
and conclusions}
\label{summary}

We analyzed the \nustar\ observation of the center of the Coma Cluster.
The main results of our work can be summarized as follows:
\begin{itemize}
\item The \nustar\ spectrum of the Coma center extracted from a 12\arcmin $\times$ 12\arcmin\
region is consistent with the superposition of thermal components in the range 7--10 keV.
The 90\% upper limit on the presence of the IC component is $5.1 \times 10^{-12}$ \fxunits.
This number is not as stringent as the one derived from modern non-imaging instruments
such as \suzaku\ HXD-PIN or \swift-BAT due to the limited FOV of \nustar\ and the high brightness of the
thermal component in the very center of the Coma Cluster.
\item The \nustar\ temperature map is in good agreement with previous measurements, in particular the one obtained
with recent deep \chandra\ exposures, and it highlights the potential of \nustar\ in constraining hot thermal plasma
in galaxy clusters.
\end{itemize}

Future mosaic \nustar\ observations of the Coma Cluster are planned to extend the coverage
to the central 30\arcmin\ $\times$ 30\arcmin\ area with 16 partially overlapping pointings
with similar exposure time to the observation reported in this work.
The Coma mosaic will constitute an excellent
legacy dataset and will be able to address some of the unresolved questions
related to this ``old friend'' cluster.\\

\begin{acknowledgements}
This research made use of data from the \nustar\ mission, a project led by the 
California Institute of Technology, managed by the Jet Propulsion Laboratory, 
and funded by NASA. 
We thank the \nustar\ Operations, Software and Calibration teams for support 
with the execution and analysis of these observations. 
This research has made use of the \nustar\ Data Analysis Software 
(NuSTARDAS) jointly developed by the ASI Science Data Center (ASDC, Italy) 
and the California Institute of Technology (USA).
\end{acknowledgements}
{\it Facility:} \facility{NuSTAR}
%
%
\bibliographystyle{apj}
\bibliography{gasta}

\begin{thebibliography}{53}
\expandafter\ifx\csname natexlab\endcsname\relax\def\natexlab#1{#1}\fi

\bibitem[{{Adami} {et~al.}(2005){Adami}, {Biviano}, {Durret}, \&
  {Mazure}}]{Adami.ea:05}
{Adami}, C., {Biviano}, A., {Durret}, F., \& {Mazure}, A. 2005, \aap, 443, 17

\bibitem[{{Adami} {et~al.}(2009){Adami}, {Le Brun}, {Biviano}, {Durret},
  {Lamareille}, {Pell{\'o}}, {Ilbert}, {Mazure}, {Trilling}, \&
  {Ulmer}}]{Adami.ea:09}
{Adami}, C., {Le Brun}, V., {Biviano}, A., {Durret}, F., {Lamareille}, F.,
  {Pell{\'o}}, R., {Ilbert}, O., {Mazure}, A., {Trilling}, R., \& {Ulmer},
  M.~P. 2009, \aap, 507, 1225

\bibitem[{{Ajello} {et~al.}(2009){Ajello}, {Rebusco}, {Cappelluti}, {Reimer},
  {B{\"o}hringer}, {Greiner}, {Gehrels}, {Tueller}, \&
  {Moretti}}]{Ajello.ea:09}
{Ajello}, M., {Rebusco}, P., {Cappelluti}, N., {Reimer}, O., {B{\"o}hringer},
  H., {Greiner}, J., {Gehrels}, N., {Tueller}, J., \& {Moretti}, A. 2009, \apj,
  690, 367

\bibitem[{{Anders} \& {Grevesse}(1989)}]{Anders.ea:89}
{Anders}, E. \& {Grevesse}, N. 1989, \gca, 53, 197

\bibitem[{{Andrade-Santos} {et~al.}(2013){Andrade-Santos}, {Nulsen}, {Kraft},
  {Forman}, {Jones}, {Churazov}, \& {Vikhlinin}}]{Andrade-Santos.ea:13}
{Andrade-Santos}, F., {Nulsen}, P.~E.~J., {Kraft}, R.~P., {Forman}, W.~R.,
  {Jones}, C., {Churazov}, E., \& {Vikhlinin}, A. 2013, \apj, 766, 107

\bibitem[{{Arnaud}(1996)}]{Arnaud:96}
{Arnaud}, K.~A. 1996, in ASP Conf. Ser. 101: Astronomical Data Analysis
  Software and Systems V, Vol.~5, 17

\bibitem[{{Arnaud} {et~al.}(2001){Arnaud}, {Neumann}, {Aghanim}, {Gastaud},
  {Majerowicz}, \& {Hughes}}]{Arnaud.ea:01}
{Arnaud}, M., {Neumann}, D.~M., {Aghanim}, N., {Gastaud}, R., {Majerowicz}, S.,
  \& {Hughes}, J.~P. 2001, \aap, 365, L80

\bibitem[{{Biviano}(1998)}]{Biviano:98}
{Biviano}, A. 1998, in Untangling Coma Berenices: A New Vision of an Old
  Cluster, ed. A.~{Mazure}, F.~{Casoli}, F.~{Durret}, \& D.~{Gerbal}, 1

\bibitem[{{Briel} {et~al.}(1992){Briel}, {Henry}, \&
  {Boehringer}}]{Briel.ea:92}
{Briel}, U.~G., {Henry}, J.~P., \& {Boehringer}, H. 1992, \aap, 259, L31

\bibitem[{{Brown} \& {Rudnick}(2011)}]{Brown.ea:11}
{Brown}, S. \& {Rudnick}, L. 2011, \mnras, 412, 2

\bibitem[{{Cavagnolo} {et~al.}(2009){Cavagnolo}, {Donahue}, {Voit}, \&
  {Sun}}]{Cavagnolo.ea:09}
{Cavagnolo}, K.~W., {Donahue}, M., {Voit}, G.~M., \& {Sun}, M. 2009, \apjs,
  182, 12

\bibitem[{{Colless} \& {Dunn}(1996)}]{Colless.ea:96}
{Colless}, M. \& {Dunn}, A.~M. 1996, \apj, 458, 435

\bibitem[{{Deiss} {et~al.}(1997){Deiss}, {Reich}, {Lesch}, \&
  {Wielebinski}}]{Deiss.ea:97}
{Deiss}, B.~M., {Reich}, W., {Lesch}, H., \& {Wielebinski}, R. 1997, \aap, 321,
  55

\bibitem[{{Donnelly} {et~al.}(1999){Donnelly}, {Markevitch}, {Forman}, {Jones},
  {Churazov}, \& {Gilfanov}}]{Donnelly.ea:99}
{Donnelly}, R.~H., {Markevitch}, M., {Forman}, W., {Jones}, C., {Churazov}, E.,
  \& {Gilfanov}, M. 1999, \apj, 513, 690

\bibitem[{{Feretti} {et~al.}(2012){Feretti}, {Giovannini}, {Govoni}, \&
  {Murgia}}]{Feretti.ea:12}
{Feretti}, L., {Giovannini}, G., {Govoni}, F., \& {Murgia}, M. 2012, \aapr, 20,
  54

\bibitem[{{Finoguenov} {et~al.}(2004){Finoguenov}, {Briel}, {Henry}, {Gavazzi},
  {Iglesias-Paramo}, \& {Boselli}}]{Finoguenov.ea:04}
{Finoguenov}, A., {Briel}, U.~G., {Henry}, J.~P., {Gavazzi}, G.,
  {Iglesias-Paramo}, J., \& {Boselli}, A. 2004, \aap, 419, 47

\bibitem[{{Fusco-Femiano} {et~al.}(1999){Fusco-Femiano}, {dal Fiume},
  {Feretti}, {Giovannini}, {Grandi}, {Matt}, {Molendi}, \&
  {Santangelo}}]{Fusco-Femiano.ea:99}
{Fusco-Femiano}, R., {dal Fiume}, D., {Feretti}, L., {Giovannini}, G.,
  {Grandi}, P., {Matt}, G., {Molendi}, S., \& {Santangelo}, A. 1999, \apjl,
  513, L21

\bibitem[{{Fusco-Femiano} {et~al.}(2007){Fusco-Femiano}, {Landi}, \&
  {Orlandini}}]{Fusco-Femiano.ea:07}
{Fusco-Femiano}, R., {Landi}, R., \& {Orlandini}, M. 2007, \apjl, 654, L9

\bibitem[{{Fusco-Femiano} {et~al.}(2011){Fusco-Femiano}, {Orlandini},
  {Bonamente}, \& {Lapi}}]{Fusco-Femiano.ea:11}
{Fusco-Femiano}, R., {Orlandini}, M., {Bonamente}, M., \& {Lapi}, A. 2011,
  \apj, 732, 85

\bibitem[{{Fusco-Femiano} {et~al.}(2004){Fusco-Femiano}, {Orlandini},
  {Brunetti}, {Feretti}, {Giovannini}, {Grandi}, \&
  {Setti}}]{Fusco-Femiano.ea:04}
{Fusco-Femiano}, R., {Orlandini}, M., {Brunetti}, G., {Feretti}, L.,
  {Giovannini}, G., {Grandi}, P., \& {Setti}, G. 2004, \apjl, 602, L73

\bibitem[{{Gavazzi} {et~al.}(2009){Gavazzi}, {Adami}, {Durret}, {Cuillandre},
  {Ilbert}, {Mazure}, {Pell{\'o}}, \& {Ulmer}}]{Gavazzi.ea:09}
{Gavazzi}, R., {Adami}, C., {Durret}, F., {Cuillandre}, J.-C., {Ilbert}, O.,
  {Mazure}, A., {Pell{\'o}}, R., \& {Ulmer}, M.~P. 2009, \aap, 498, L33

\bibitem[{{Giovannini} {et~al.}(1993){Giovannini}, {Feretti}, {Venturi}, {Kim},
  \& {Kronberg}}]{Giovannini.ea:93}
{Giovannini}, G., {Feretti}, L., {Venturi}, T., {Kim}, K.-T., \& {Kronberg},
  P.~P. 1993, \apj, 406, 399

\bibitem[{{Goodman} \& {Weare}(2010)}]{Goodman.ea:10}
{Goodman}, J. \& {Weare}, J. 2010, Comm. App. Math.~Comp.~Sci., 5, 65

\bibitem[{{Harris} \& {Romanishin}(1974)}]{Harris.ea:74}
{Harris}, D.~E. \& {Romanishin}, W. 1974, \apj, 188, 209

\bibitem[{{Harrison} {et~al.}(2013){Harrison}, {Craig}, {Christensen},
  {Hailey}, {Zhang}, {Boggs}, {Stern}, {Cook}, {Forster}, {Giommi},
  {Grefenstette}, {Kim}, {Kitaguchi}, {Koglin}, {Madsen}, {Mao}, {Miyasaka},
  {Mori}, {Perri}, {Pivovaroff}, {Puccetti}, {Rana}, {Westergaard}, {Willis},
  {Zoglauer}, {An}, {Bachetti}, {Barri{\`e}re}, {Bellm}, {Bhalerao},
  {Brejnholt}, {Fuerst}, {Liebe}, {Markwardt}, {Nynka}, {Vogel}, {Walton},
  {Wik}, {Alexander}, {Cominsky}, {Hornschemeier}, {Hornstrup}, {Kaspi},
  {Madejski}, {Matt}, {Molendi}, {Smith}, {Tomsick}, {Ajello}, {Ballantyne},
  {Balokovi{\'c}}, {Barret}, {Bauer}, {Blandford}, {Brandt}, {Brenneman},
  {Chiang}, {Chakrabarty}, {Chenevez}, {Comastri}, {Dufour}, {Elvis}, {Fabian},
  {Farrah}, {Fryer}, {Gotthelf}, {Grindlay}, {Helfand}, {Krivonos}, {Meier},
  {Miller}, {Natalucci}, {Ogle}, {Ofek}, {Ptak}, {Reynolds}, {Rigby},
  {Tagliaferri}, {Thorsett}, {Treister}, \& {Urry}}]{Harrison.ea:13}
{Harrison}, F.~A., {Craig}, W.~W., {Christensen}, F.~E., {Hailey}, C.~J.,
  {Zhang}, W.~W., {Boggs}, S.~E., {Stern}, D., {Cook}, W.~R., {Forster}, K.,
  {Giommi}, P., {Grefenstette}, B.~W., {Kim}, Y., {Kitaguchi}, T., {Koglin},
  J.~E., {Madsen}, K.~K., {Mao}, P.~H., {Miyasaka}, H., {Mori}, K., {Perri},
  M., {Pivovaroff}, M.~J., {Puccetti}, S., {Rana}, V.~R., {Westergaard}, N.~J.,
  {Willis}, J., {Zoglauer}, A., {An}, H., {Bachetti}, M., {Barri{\`e}re},
  N.~M., {Bellm}, E.~C., {Bhalerao}, V., {Brejnholt}, N.~F., {Fuerst}, F.,
  {Liebe}, C.~C., {Markwardt}, C.~B., {Nynka}, M., {Vogel}, J.~K., {Walton},
  D.~J., {Wik}, D.~R., {Alexander}, D.~M., {Cominsky}, L.~R., {Hornschemeier},
  A.~E., {Hornstrup}, A., {Kaspi}, V.~M., {Madejski}, G.~M., {Matt}, G.,
  {Molendi}, S., {Smith}, D.~M., {Tomsick}, J.~A., {Ajello}, M., {Ballantyne},
  D.~R., {Balokovi{\'c}}, M., {Barret}, D., {Bauer}, F.~E., {Blandford}, R.~D.,
  {Brandt}, W.~N., {Brenneman}, L.~W., {Chiang}, J., {Chakrabarty}, D.,
  {Chenevez}, J., {Comastri}, A., {Dufour}, F., {Elvis}, M., {Fabian}, A.~C.,
  {Farrah}, D., {Fryer}, C.~L., {Gotthelf}, E.~V., {Grindlay}, J.~E.,
  {Helfand}, D.~J., {Krivonos}, R., {Meier}, D.~L., {Miller}, J.~M.,
  {Natalucci}, L., {Ogle}, P., {Ofek}, E.~O., {Ptak}, A., {Reynolds}, S.~P.,
  {Rigby}, J.~R., {Tagliaferri}, G., {Thorsett}, S.~E., {Treister}, E., \&
  {Urry}, C.~M. 2013, \apj, 770, 103

\bibitem[{{Kalberla} {et~al.}(2005){Kalberla}, {Burton}, {Hartmann}, {Arnal},
  {Bajaja}, {Morras}, \& {P{\"o}ppel}}]{Kalberla.ea:05}
{Kalberla}, P.~M.~W., {Burton}, W.~B., {Hartmann}, D., {Arnal}, E.~M.,
  {Bajaja}, E., {Morras}, R., \& {P{\"o}ppel}, W.~G.~L. 2005, \aap, 440, 775

\bibitem[{{Mazzotta} {et~al.}(2004){Mazzotta}, {Rasia}, {Moscardini}, \&
  {Tormen}}]{Mazzotta.ea:04}
{Mazzotta}, P., {Rasia}, E., {Moscardini}, L., \& {Tormen}, G. 2004, \mnras,
  354, 10

\bibitem[{{Neumann} {et~al.}(2003){Neumann}, {Lumb}, {Pratt}, \&
  {Briel}}]{Neumann.ea:03}
{Neumann}, D.~M., {Lumb}, D.~H., {Pratt}, G.~W., \& {Briel}, U.~G. 2003, \aap,
  400, 811

\bibitem[{{Okabe} {et~al.}(2014){Okabe}, {Futamase}, {Kajisawa}, \&
  {Kuroshima}}]{Okabe.ea:14}
{Okabe}, N., {Futamase}, T., {Kajisawa}, M., \& {Kuroshima}, R. 2014, \apj,
  784, 90

\bibitem[{{Okabe} {et~al.}(2010){Okabe}, {Okura}, \& {Futamase}}]{Okabe.ea:10}
{Okabe}, N., {Okura}, Y., \& {Futamase}, T. 2010, \apj, 713, 291

\bibitem[{{Ota}(2012)}]{Ota:12}
{Ota}, N. 2012, Research in Astronomy and Astrophysics, 12, 973

\bibitem[{{Planck Collaboration} {et~al.}(2013){Planck Collaboration}, {Ade},
  {Aghanim}, {Arnaud}, {Ashdown}, {Atrio-Barandela}, {Aumont}, {Baccigalupi},
  {Balbi}, {Banday}, \& et~al.}]{PlanckIntermediateX:13}
{Planck Collaboration}, {Ade}, P.~A.~R., {Aghanim}, N., {Arnaud}, M.,
  {Ashdown}, M., {Atrio-Barandela}, F., {Aumont}, J., {Baccigalupi}, C.,
  {Balbi}, A., {Banday}, A.~J., \& et~al. 2013, \aap, 554, A140

\bibitem[{{Rephaeli}(1977)}]{Rephaeli:77}
{Rephaeli}, Y. 1977, \apj, 212, 608

\bibitem[{{Rephaeli} \& {Gruber}(2002)}]{Rephaeli.ea:02}
{Rephaeli}, Y. \& {Gruber}, D. 2002, \apj, 579, 587

\bibitem[{{Rephaeli} {et~al.}(1999){Rephaeli}, {Gruber}, \&
  {Blanco}}]{Rephaeli.ea:99}
{Rephaeli}, Y., {Gruber}, D., \& {Blanco}, P. 1999, \apjl, 511, L21

\bibitem[{{Rephaeli} {et~al.}(2008){Rephaeli}, {Nevalainen}, {Ohashi}, \&
  {Bykov}}]{Rephaeli.ea:08}
{Rephaeli}, Y., {Nevalainen}, J., {Ohashi}, T., \& {Bykov}, A.~M. 2008, \ssr,
  134, 71

\bibitem[{{Rossetti} {et~al.}(2007){Rossetti}, {Ghizzardi}, {Molendi}, \&
  {Finoguenov}}]{Rossetti.ea:07}
{Rossetti}, M., {Ghizzardi}, S., {Molendi}, S., \& {Finoguenov}, A. 2007, \aap,
  463, 839

\bibitem[{{Rossetti} \& {Molendi}(2004)}]{Rossetti.ea:04}
{Rossetti}, M. \& {Molendi}, S. 2004, \aap, 414, L41

\bibitem[{{Sanders} {et~al.}(2013){Sanders}, {Fabian}, {Churazov},
  {Schekochihin}, {Simionescu}, {Walker}, \& {Werner}}]{Sanders.ea:13b}
{Sanders}, J.~S., {Fabian}, A.~C., {Churazov}, E., {Schekochihin}, A.~A.,
  {Simionescu}, A., {Walker}, S.~A., \& {Werner}, N. 2013, Science, 341, 1365

\bibitem[{{Sato} {et~al.}(2011){Sato}, {Matsushita}, {Ota}, {Sato}, {Nakazawa},
  \& {Sarazin}}]{Sato.ea:11}
{Sato}, T., {Matsushita}, K., {Ota}, N., {Sato}, K., {Nakazawa}, K., \&
  {Sarazin}, C.~L. 2011, \pasj, 63, 991

\bibitem[{{Schuecker} {et~al.}(2004){Schuecker}, {Finoguenov}, {Miniati},
  {B{\"o}hringer}, \& {Briel}}]{Schuecker.ea:04}
{Schuecker}, P., {Finoguenov}, A., {Miniati}, F., {B{\"o}hringer}, H., \&
  {Briel}, U.~G. 2004, \aap, 426, 387

\bibitem[{{Simionescu} {et~al.}(2013){Simionescu}, {Werner}, {Urban}, {Allen},
  {Fabian}, {Mantz}, {Matsushita}, {Nulsen}, {Sanders}, {Sasaki}, {Sato},
  {Takei}, \& {Walker}}]{Simionescu.ea:13}
{Simionescu}, A., {Werner}, N., {Urban}, O., {Allen}, S.~W., {Fabian}, A.~C.,
  {Mantz}, A., {Matsushita}, K., {Nulsen}, P.~E.~J., {Sanders}, J.~S.,
  {Sasaki}, T., {Sato}, T., {Takei}, Y., \& {Walker}, S.~A. 2013, \apj, 775, 4

\bibitem[{{Smith} {et~al.}(2001){Smith}, {Brickhouse}, {Liedahl}, \&
  {Raymond}}]{Smith.ea:01}
{Smith}, R.~K., {Brickhouse}, N.~S., {Liedahl}, D.~A., \& {Raymond}, J.~C.
  2001, \apjl, 556, L91

\bibitem[{{Sun} {et~al.}(2007){Sun}, {Jones}, {Forman}, {Vikhlinin}, {Donahue},
  \& {Voit}}]{Sun.ea:07}
{Sun}, M., {Jones}, C., {Forman}, W., {Vikhlinin}, A., {Donahue}, M., \&
  {Voit}, M. 2007, \apj, 657, 197

\bibitem[{{Thierbach} {et~al.}(2003){Thierbach}, {Klein}, \&
  {Wielebinski}}]{Thierbach.ea:03}
{Thierbach}, M., {Klein}, U., \& {Wielebinski}, R. 2003, \aap, 397, 53

\bibitem[{{Vikhlinin} {et~al.}(1994){Vikhlinin}, {Forman}, \&
  {Jones}}]{Vikhlinin.ea:94}
{Vikhlinin}, A., {Forman}, W., \& {Jones}, C. 1994, \apj, 435, 162

\bibitem[{{Vikhlinin} {et~al.}(1997){Vikhlinin}, {Forman}, \&
  {Jones}}]{Vikhlinin.ea:97}
---. 1997, \apjl, 474, L7

\bibitem[{{Vikhlinin} {et~al.}(1999){Vikhlinin}, {Forman}, \&
  {Jones}}]{vikhlinin.ea:99}
---. 1999, \apj, 525, 47

\bibitem[{{Watanabe} {et~al.}(1999){Watanabe}, {Yamashita}, {Furuzawa},
  {Kunieda}, {Tawara}, \& {Honda}}]{Watanabe.ea:99}
{Watanabe}, M., {Yamashita}, K., {Furuzawa}, A., {Kunieda}, H., {Tawara}, Y.,
  \& {Honda}, H. 1999, \apj, 527, 80

\bibitem[{{Wik} {et~al.}(2014){Wik}, {Hornstrup}, {Molendi}, {Madejski},
  {Harrison}, {Zoglauer}, {Grefenstette}, {Gastaldello}, {Madsen},
  {Westergaard}, {Ferreira}, {Kitaguchi}, {Pedersen}, {Boggs}, {Christensen},
  {Craig}, {Hailey}, {Stern}, \& {Zhang}}]{Wik.ea:14}
{Wik}, D.~R., {Hornstrup}, A., {Molendi}, S., {Madejski}, G., {Harrison},
  F.~A., {Zoglauer}, A., {Grefenstette}, B.~W., {Gastaldello}, F., {Madsen},
  K.~K., {Westergaard}, N.~J., {Ferreira}, D.~D.~M., {Kitaguchi}, T.,
  {Pedersen}, K., {Boggs}, S.~E., {Christensen}, F.~E., {Craig}, W.~W.,
  {Hailey}, C.~J., {Stern}, D., \& {Zhang}, W.~W. 2014, \apj, 792, 48

\bibitem[{{Wik} {et~al.}(2011){Wik}, {Sarazin}, {Finoguenov}, {Baumgartner},
  {Mushotzky}, {Okajima}, {Tueller}, \& {Clarke}}]{Wik.ea:11}
{Wik}, D.~R., {Sarazin}, C.~L., {Finoguenov}, A., {Baumgartner}, W.~H.,
  {Mushotzky}, R.~F., {Okajima}, T., {Tueller}, J., \& {Clarke}, T.~E. 2011,
  \apj, 727, 119

\bibitem[{{Wik} {et~al.}(2009){Wik}, {Sarazin}, {Finoguenov}, {Matsushita},
  {Nakazawa}, \& {Clarke}}]{Wik.ea:09}
{Wik}, D.~R., {Sarazin}, C.~L., {Finoguenov}, A., {Matsushita}, K., {Nakazawa},
  K., \& {Clarke}, T.~E. 2009, \apj, 696, 1700

\bibitem[{{Willson}(1970)}]{Willson.ea:70}
{Willson}, M.~A.~G. 1970, \mnras, 151, 1

\end{thebibliography}

\end{document}